\documentclass[letterpaper]{article} 

\usepackage{aaai2026}  
\usepackage{times}  
\usepackage{helvet}  
\usepackage{courier}  
\usepackage[hyphens]{url}  
\usepackage{graphicx} 
\usepackage{natbib}  
\usepackage{caption} 
\usepackage{algorithm}
\usepackage{newfloat}
\usepackage{listings}
\DeclareCaptionStyle{ruled}{labelfont=normalfont,labelsep=colon,strut=off} 
\floatstyle{ruled}
\newfloat{listing}{tb}{lst}{}
\floatname{listing}{Listing}

\usepackage{amsmath, amssymb, algorithm, algpseudocode, multirow, adjustbox, booktabs, makecell} 

\usepackage{microtype}
\usepackage{latexsym}
\usepackage{pifont}

\usepackage{svg}

\usepackage{booktabs}
\usepackage{array}
\usepackage{longtable}

\newcommand{\quotebox}[1]{\begin{center}\begin{minipage}{\linewidth}\vspace{5pt}\center\begin{minipage}{0.9\linewidth}{\space\Huge``}{#1}{\hspace{1.5em}\break\null\Huge\hfill''}\end{minipage}\smallbreak\end{minipage}\end{center}}

\title{Joint-GCG: Unified Gradient-Based Poisoning Attacks on Retrieval-Augmented Generation Systems}

\author{
    Haowei Wang\textsuperscript{\rm 1, 2, 3}\equalcontrib,
    Rupeng Zhang\textsuperscript{\rm 1, 2, 3}\equalcontrib,
    Junjie Wang\textsuperscript{\rm 1, 2, 3}\thanks{Corresponding authors.},
    Mingyang Li\textsuperscript{\rm 1, 2, 3},
    Yuekai Huang\textsuperscript{\rm 1, 2, 3},
    Dandan Wang\textsuperscript{\rm 1, 2, 3}\footnotemark[2],
    Qing Wang\textsuperscript{\rm 1, 2, 3}
}
\affiliations{
    \textsuperscript{\rm 1}State Key Laboratory of Complex System Modeling and Simulation Technology, Beijing, China\\
    \textsuperscript{\rm 2}Institute of Software, Chinese Academy of Sciences, Beijing, China\\
    \textsuperscript{\rm 3}University of Chinese Academy of Sciences, Beijing, China\\
    \{wanghaowei2023, zhangrupeng2023\}@iscas.ac.cn\\
    
    \{junjie, dandan\}@iscas.ac.cn
}

\begin{document}

\maketitle

\begin{abstract}
Retrieval-Augmented Generation (RAG) systems enhance Large Language Models (LLMs) by retrieving relevant documents from external corpora before generating responses. This approach significantly expands LLM capabilities by leveraging vast, up-to-date external knowledge. However, this reliance on external knowledge makes RAG systems vulnerable to corpus poisoning attacks that manipulate generated outputs via poisoned document injection.
Existing poisoning attack strategies typically treat the retrieval and generation stages as disjointed, limiting their effectiveness. 
We propose \textbf{Joint-GCG}, the first framework to unify gradient-based attacks across both retriever and generator models through three innovations: (1) \textit{Cross-Vocabulary Projection} for aligning embedding spaces, (2) \textit{Gradient Tokenization Alignment} for synchronizing token-level gradient signals, and (3) \textit{Adaptive Weighted Fusion} for dynamically balancing attacking objectives. Evaluations demonstrate that Joint-GCG achieves at most 25\% and an average of 5\% higher attack success rate than previous methods across multiple retrievers and generators. 
While optimized under a white-box assumption, the generated poisons show unprecedented transferability to unseen models.
Joint-GCG's innovative unification of gradient-based attacks across retrieval and generation stages fundamentally reshapes our understanding of vulnerabilities within RAG systems.
\end{abstract}

\begin{links}
    \link{Code}{https://github.com/NicerWang/Joint-GCG}
\end{links}

\section{Introduction}

Retrieval-Augmented Generation (RAG) systems~\cite{lewis2020retrieval, ram2023context} have emerged as a powerful paradigm for enhancing Large Language Models (LLMs). By coupling a retriever, which fetches relevant documents from an external corpus based on a given query, and a generator that synthesizes information to produce coherent and contextually appropriate responses, RAG systems leverage vast, up-to-date external knowledge. This architecture significantly improves the performance of diverse AI applications—including search engines~\cite{chatgptsearch}, chatbots~\cite{vakayil2024rag, boulos2024nvidia}, code assistants~\cite{zhou2022docprompting, nashid2023retrieval}, and knowledge bases~\cite{siriwardhana2023improving, meduri2024efficient}—by ensuring outputs are accurate and up-to-date~\cite{fan2024survey}.

\begin{figure}[ht]
    \centering
    \includegraphics[width=\linewidth]{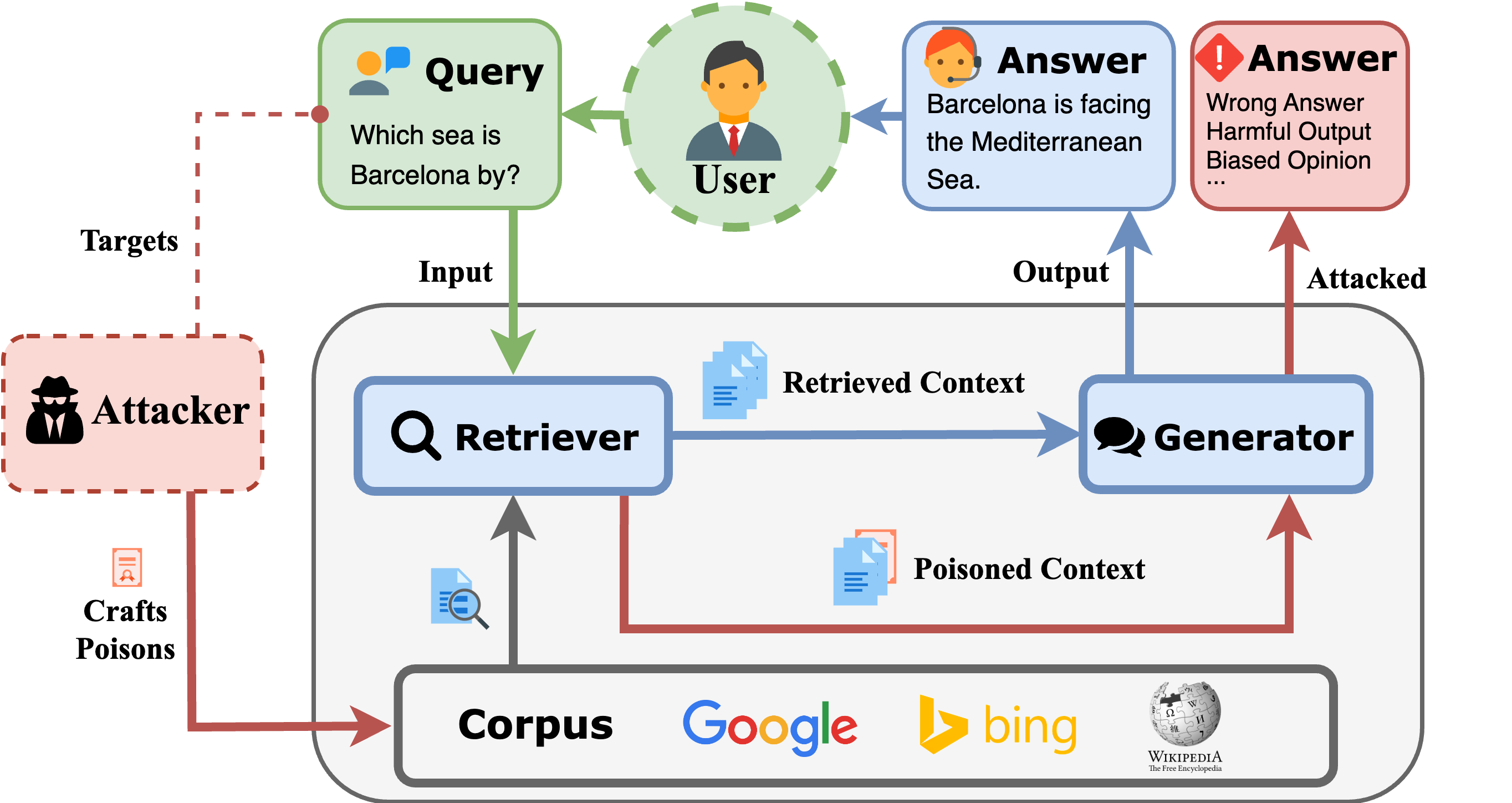}
    \caption{Demonstration of RAG systems and RAG-poisoning attacks. We provide an example of a successful attack on RAG systems to induce a wrong answer in Appendix~\ref{A-tab: poisoned-example}.}
    \label{fig:rag-poisoning}
\end{figure}

However, this remarkable power comes with a critical vulnerability: reliance on external corpora introduces the risk of corpus poisoning~\cite{zou2024poisonedrag, xue2024badrag, tan2024glue, cheng2024trojanrag, chaudhari2024phantom}. As illustrated in Figure~\ref{fig:rag-poisoning}, corpus poisoning involves malicious actors injecting crafted poisoned documents into the knowledge base. If retrieved and processed, these poisoned documents can cause the RAG system to generate wrong answers, harmful outputs, or biased opinions, thereby undermining its reliability. 
Besides, the growing trend of RAG systems employing open-source components is intended to facilitate transparency, customization, and data leakage prevention. However, this allows attackers to study and replicate system architectures meticulously, making RAG systems more susceptible to corpus poisoning attacks.

The objective of corpus poisoning is to introduce poisoned documents into the corpus and ensure they can be retrieved by targeted queries. More importantly, the attacker must manipulate the subsequent generation process to ensure the model produces erroneous outputs based on the poisoned information. Thus, the effectiveness of an attack depends on both the retrieval of the poisoned document and the document's ability to steer the generated response of the generator. Crucially, attackers aim to achieve this influence for stealth and practicality by injecting as few poisoned documents as possible, which is essential for evading detection.

Existing attack strategies, as detailed in the Appendix~\ref{A-related-works}, often adopt a fragmented approach, treating the retrieval and generation stages as disjoint optimization problems.
For instance, Phantom~\cite{chaudhari2024phantom} and LIAR~\cite{tan2024glue} tackle the retriever and generator objectives independently and sequentially.
Such methods can be suboptimal, as they overlook the synergistic effects that could be achieved by simultaneously optimizing for both components, potentially limiting the overall efficacy of the poisoning attack.

To address this limitation, we propose \textbf{Joint-GCG}, a novel framework that, for the first time, unifies the attack surface by jointly optimizing gradients and losses across both the retriever and the generator.
Joint-GCG overcomes the technical hurdles of joint optimization, including mismatched vocabularies and differing tokenization schemes between models, through three key innovations: (1) \textit{Cross-Vocabulary Projection (CVP)}, which aligns vocabulary embeddings; (2) \textit{Gradient Tokenization Alignment (GTA)}, which synchronizes token-level gradient signals; and (3) \textit{Adaptive Weighted Fusion (AWF)}, which dynamically balances the influence of retrieval and generation objectives. These innovative mechanisms work together to form a highly effective attack strategy, showcasing the power of joint optimization in RAG poisoning.

Our key contributions are:
\begin{itemize}
    \item \textbf{Problem Modeling Innovation:} We identify the limitations of existing disjointed attack strategies and highlight the critical need for joint optimization of retrieval and generation in RAG poisoning. We are the first to emphasize the synergistic potential of unified optimization for significantly enhanced attack efficacy, shifting the paradigm from independent component attacks to a holistic system-level approach.

    \item \textbf{Novel Joint Optimization Framework:} We propose Joint-GCG, a novel framework that effectively addresses the challenges of joint optimization in RAG poisoning, enabling more effective and accurately guided corpus poisoning by orchestrating a synergistic attack across the retrieval and generation stages.
    
    \item \textbf{Systematic Evaluation:} We demonstrate Joint-GCG's superiority over state-of-the-art methods with at most 25\% and an average of 5\% higher attack success rate on average, achieves significant cross-retriever transferability as well as notable cross-generator transferability (achieving at most 57\% $ASR$ on unseen models), and showcase its applicability in batch poisoning and synthetic corpus scenarios, amplifying the vulnerability of RAG systems.
\end{itemize}

\section{Threat Model}
Our threat model accounts for the capabilities and knowledge assumed by the attacker when crafting poisons for RAG systems using Joint-GCG. It is designed to facilitate a thorough investigation of potential vulnerabilities, particularly those arising from the joint optimization of retriever and generator components.

\textbf{White-box Retriever and Generator Access:} We assume the attacker has \textit{full white-box access} to both the retriever and the generator models. This comprehensive accessibility means the attacker possesses knowledge of model architectures, including all layers and configurations, has access to all model parameters, and can compute gradients of any loss function concerning model inputs (i.e., the optimizable poison sequence).

This white-box assumption is increasingly pertinent given the proliferation of open-source RAG components (retrievers and LLMs) that attackers can replicate or directly access, making systems built with these components vulnerable to this level of scrutiny.

Furthermore, understanding system vulnerabilities under complete information is often a prerequisite for developing robust defenses, and insights from white-box attacks can also guide the creation of more practical gray-box or black-box attack strategies. 

Finally, this approach is consistent with methodologies in several contemporary RAG poisoning research works (e.g., LIAR~\cite{tan2024glue}, Phantom~\cite{chaudhari2024phantom}), which also operate under white-box assumptions for both components, establishing this as a standard paradigm for in-depth analysis in this research area. The strong poison transferability demonstrated in our experiments reveals a practical path for relaxing this white-box assumption during attack deployment. An attacker can leverage Joint-GCG in a white-box setting using a powerful, locally hosted surrogate model that mimics the target's possible architecture. The resulting poison can then be utilized to attack the target system, even if the components are different or entirely black-box. Our results show that poisons optimized on one generator can successfully attack others. Similarly, poisons demonstrate high transferability across different retriever architectures. This surrogate model approach transforms the attack into a gray-box scenario, where the attacker only needs to inject the pre-crafted document into the target corpus without internal access to the production models.

\textbf{Gray-box Corpus Access:} We assume the attacker has limited, or \textit{gray-box}, access to the retrieval corpus. In our experiments, the attacker can inject a small, fixed number of poisoned documents into the corpus, typically one poisoned document per target query to ensure stealth, but cannot modify or delete existing legitimate documents. Our Adaptive Weighted Fusion (AWF) module calculates a stability metric by analyzing the top-$k$ documents retrieved for a query. Given that RAG systems often cite their sources, effectively making the top retrieved documents accessible, we find it realistic for the attacker to observe these results during the optimization phase. We also explore scenarios using synthetic corpora to mitigate this specific observation requirement. This gray-box corpus access reflects realistic scenarios such as decentralized knowledge bases, wikis, or systems that index publicly editable web content, where attackers can introduce malicious content but do not have control over the entire corpus. The constraint on the number of injected documents reflects the practical need for stealth, as injecting a large volume of suspicious documents would likely be detected.

This comprehensive threat model enables Joint-GCG to examine the intricate relationships between the retriever and generator during a poisoning attack, offering in-depth insights into the security posture of modern RAG systems.

\begin{figure*}[t]
    \centering
    \includegraphics[width=0.9\linewidth]{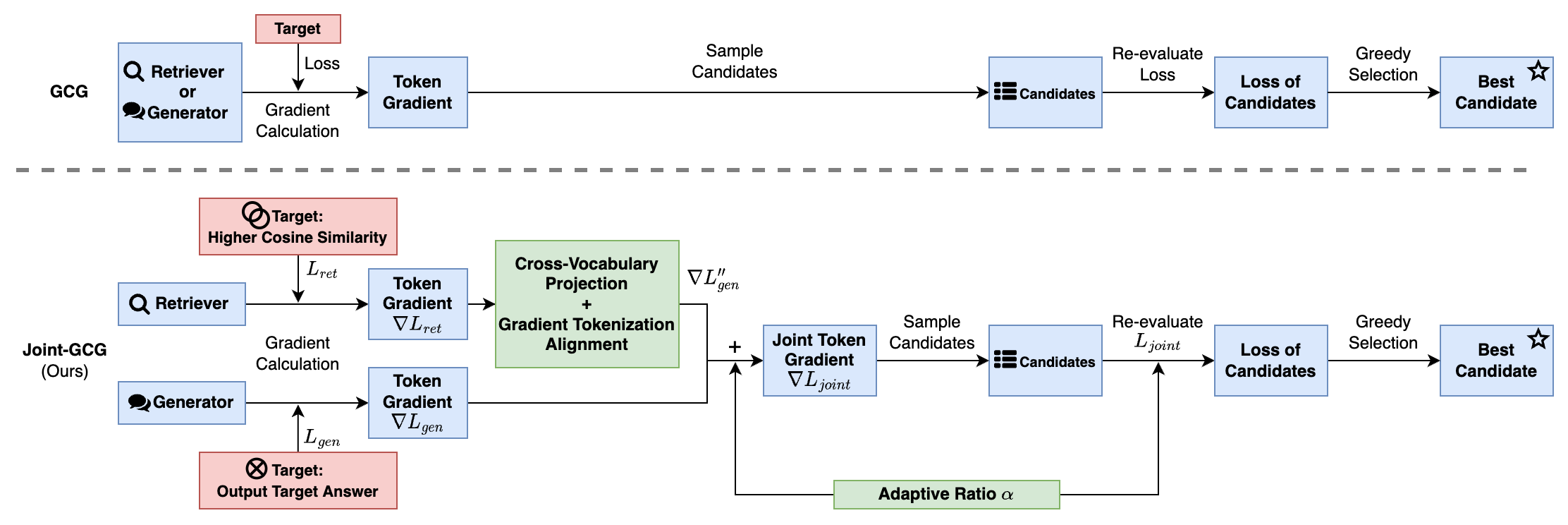}
    \caption{The optimizing process of the Joint-GCG framework, compared to regular GCG.}
    \label{fig: joint-gcg}
\end{figure*}

\section{Methodology: Joint-GCG Framework}
To address the limitations in disjointed RAG poisoning attacks, we propose \textbf{Joint-GCG}, a novel framework designed to unify the attack process by simultaneously targeting both the retriever and the generator.

Inspired by the success of Greedy Coordinate Gradient (GCG) techniques~\cite{zou2023universal} in manipulating generator outputs through gradient-guided input optimization, \textbf{Joint-GCG} conceptualizes the RAG system as a single, integrated target for the poisoning attack, as illustrated in Figure~\ref{fig: joint-gcg}. It adapts the iterative GCG process to optimize for a joint objective that harmonizes two distinct attack vectors: manipulating the generator's output and ensuring the poisoned document is retrieved.

To achieve this, we first define \textbf{Joint Optimization Objective}. This objective is minimized using an iterative optimization loop, which is similar to GCG. The primary challenge lies in combining gradients and losses from two architecturally disparate models. \textbf{Joint-GCG} addresses this through three key innovations: (1) \textit{Cross-Vocabulary Projection (CVP)}, which aligns the embedding spaces of disparate vocabularies; (2) \textit{Gradient Tokenization Alignment (GTA)}, which synchronizes token-level gradient signals across different tokenization outputs; and (3) \textit{Adaptive Weighted Fusion (AWF)}, which dynamically balances the attacking objectives for the retriever and generator. 

\subsection{The Joint Optimization Objective}
\begin{figure}
    \centering
    \includegraphics[width=0.7\columnwidth]{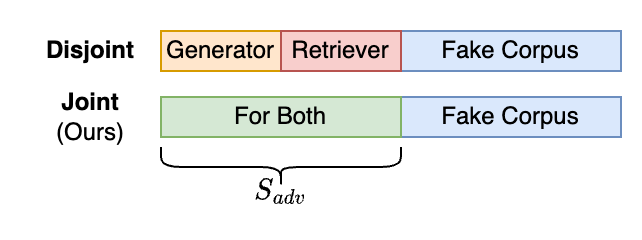}
    \caption{The $S_\text{adv}$ pattern for constructing the poisoned corpus.}
    \label{fig:sequence}
\end{figure}

As demonstrated in Figure~\ref{fig:sequence}, the attack is framed as an optimization problem to find an optimal adversarial sequence $S_\text{adv}$ that, when injected into a document to form a poisoned document $d_p$, minimizes a joint loss function $L_\text{joint}$ for a given target query $Q$.

\begin{equation}
\min_{S_\text{adv}} L_\text{joint}(S_\text{adv}; Q) = (1 - \alpha) \cdot L_\text{gen} + \alpha \cdot L_\text{ret}
\label{eq:obj}
\end{equation}

The weighting parameter $\alpha \in [0, 1]$ is determined dynamically by AWF introduced in Section~\ref{method-awd} to balance these two objectives. 

Let $\mathcal{D}_\text{ret} = TopK_{d_i\in \mathcal{D}}(E(Q), E(d_i))$ be the set of Top-K documents retrieved from the corpus $\mathcal{D}$ for the query $Q$, based on retriever model $E$. Our loss function is defined as:

\begin{equation}
L_\text{joint} = (1 - \alpha) \cdot \mathbb{I}(d_p \in \mathcal{D}_\text{ret}) \cdot L_\text{gen}  + \alpha \cdot L_\text{ret}
\label{eq:joint_loss}
\end{equation}

 $\mathbb{I}(\cdot)$ is an indicator function, which is 1 if $d_p$ is successfully retrieved into the Top-K context, and 0 otherwise. This formulation ensures that the generation loss is only considered when the poisoned document can actually influence the generator. The two loss functions are defined as:

\begin{itemize}
    \item Retrieval Loss ($L_\text{ret}$). To ensure $d_p$ is retrieved, we maximize its cosine similarity with the query $Q$ by minimizing its negative value.
\begin{equation}
L_\text{ret} = - \mathrm{sim}(Q, d_p) = - \frac{E(Q) \cdot E(d_p)}{\|E(Q)\| \|E(d_p)\|}
\end{equation}
    \item Generation Loss ($L_\text{gen}$). To guide the generator towards a target answer $T$, we use the cross-entropy loss to measure the difficulty of generating $T$ given the actual retrieved context $\mathcal{D}_{retrieved}$.
    
\begin{equation}
L_\text{gen} = - \log P(T | Q, \mathcal{D}_\text{ret}; \theta_\text{gen})
\end{equation}
\end{itemize}
\subsection{The Joint-GCG Attack Loop}
The entire Joint-GCG attack loop is designed to find a $S_\text{adv}$ that minimizes $L_\text{joint}$.
We now detail how each step of GCG algorithm is adapted for our joint attack task.

\subsubsection{Step 1: Joint Gradient Computation and Fusion}
The first step is to compute the gradient of $L_\text{joint}$ with respect to the input tokens, which requires computing the gradients from the generator ($\nabla L_\text{gen}$) and the retriever ($\nabla L_\text{ret}$). However, these raw gradients are incompatible due to differences in model vocabularies and tokenization schemes.

To resolve this, we employ Cross-Vocabulary Projection and Gradient Tokenization Alignment:
\paragraph{\textbf{Cross-Vocabulary Projection (CVP): Bridging Vocabulary Discrepancies}}
Typically, the retriever's vocabulary differs from the generator's vocabulary due to their distinct training process. CVP addresses this gap by utilizing a combination of the generator tokens to represent the retriever tokens in the embedding space.

Let $N_\text{gen}, V_\text{gen}$ represent the token number of $S_\text{adv}$ and the vocabulary size for the generator, and $N_\text{ret}, V_\text{ret}$ for the retriever. During the attack process, we compute gradient matrices $\nabla L_\text{gen} \in \mathbb{R}^{N_\text{gen} \times V_\text{gen}}$ and $\nabla L_\text{ret} \in \mathbb{R}^{N_\text{ret} \times V_\text{ret}}$.

Directly obtaining a high-dimensional linear transformation matrix $W \in \mathbb{R}^{V_{\text{ret}} \times V_{\text{gen}}}$ using generator tokens to represent retriever tokens is rendered infeasible by both the ill-posed nature of constructing a sufficient system of equations and the substantial computational demands. Instead, CVP adopts a more tractable approach by focusing on mapping individual token embeddings. Let $E_{\text{gen}} \in \mathbb{R}^{V_{\text{gen}} \times D_{\text{gen}}}$ and $E_{\text{ret}} \in \mathbb{R}^{V_{\text{ret}} \times D_{\text{ret}}}$ represent the embedding layer matrices of the generator and the retriever, respectively, where $D$ denotes the embedding dimension. For each retriever token embedding $y \in \mathbb{R}^{D_{\text{ret}}}$, our objective is to find a representation $x \in \mathbb{R}^{V_{\text{gen}}}$ such that it acts as a linear combination with the generator tokens when transformed by a learned embedding mapping function $f$, closely approximates $y$:
\begin{equation}
    f(x E_{\text{gen}}) = y
\end{equation}
Here, $f: \mathbb{R}^{D_{\text{gen}}} \to \mathbb{R}^{D_{\text{ret}}}$ represents a mapping function between the two embedding spaces.
We use the encoder of a trained autoencoder to serve as the mapping function $f$, by applying $f$ to the generator's embeddings, we can then construct and solve a system of linear equations to obtain the final projection matrix enabling $W \in \mathbb{R}^{V_{\text{ret}} \times V_{\text{gen}}}$, via concatenating all solved $x$. And then $W$ is used for transforming $\nabla L_\text{ret}$ into $\nabla L_\text{ret}' \in \mathbb{R}^{N_\text{ret} \times V_\text{gen}}$. Detailed architectural specifications of the autoencoder, training procedures, and analysis of CVP are provided in Appendix~\ref{A-CVP-additional}.

\paragraph{\textbf{Gradient Tokenization Alignment (GTA)}} After CVP, another critical challenge arises from differing tokenization schemes. Retrievers and generators often employ distinct tokenizers, resulting in differences in how $S_\text{adv}$ is segmented into tokens. To address this, we propose GTA, a module designed to synchronize gradient signals at a finer, tokenizer-agnostic granularity.

To achieve tokenization alignment, GTA employs character-level gradients as an intermediary. Retriever token gradients $\nabla L_\text{ret}'$ are decomposed and assigned to their constituent characters, recognizing characters as a more fundamental and tokenization-agnostic text unit. Subsequently, to obtain token-level gradients for the retriever, we average the gradients of the characters within each token. This averaging method robustly consolidates character-level information back into the generator's token space while mitigating potential noise from the decomposition process.

Through the GTA process, we obtain a transformed retriever gradient matrix $\nabla L_\text{gen}'' \in \mathbb{R}^{N_\text{gen} \times V_\text{gen}}$. Crucially, $\nabla L_\text{gen}''$ is now aligned with the generator's gradient matrix $\nabla L_\text{gen}$ in both sequence length and vocabulary dimensions, enabling a meaningful and direct fusion of gradient information from the retriever and the generator for joint optimization.

\paragraph{\textbf{Adaptive Weighted Fusion (AWF): Fusing the Gradients}}
\label{method-awd}
With $\nabla L_\text{gen}$ and $\nabla L_\text{gen}''$ now aligned in all dimensions, the final critical step is determining how to combine these gradient matrices to achieve joint optimization effectively. We propose \textbf{AWF}, a module that dynamically adjusts the relative contribution of each gradient matrix during the attack process. This adaptive weighting mechanism is essential because the optimal balance between prioritizing retrieval and generation objectives can vary significantly depending on the specific attack scenario, the RAG system settings, and the characteristics of the target query and the corpus.

As shown in Appendix~\ref{A-poison-pos}, we observed that poisoned documents retrieved at higher ranks are generally more influential in shaping the generator's response. To optimize attack performance, it is desirable to position the poisoned document as high as possible in the retrieval ranking while ensuring a sufficient margin from other documents to mitigate potential rank fluctuations during subsequent attack steps.

AWF introduces a stability metric $P$, quantifying the robustness of the poisoned document's retrieval rank:

\begin{equation}
    P = \frac{\mathrm{sim}(d_p, Q) - \mathrm{sim}(d_0, Q)}{D_{avg}}
\end{equation}

where $\mathrm{sim}(d_p, Q)$ and $\mathrm{sim}(d_0, Q)$ are the similarity scores of the query $Q$ with the poisoned document $d_p$ and the benign document with the highest rank, respectively, $D_{avg}$ is the average similarity score difference between consecutive Top-K documents, defined as below:

\begin{equation}
    D_{avg} = \frac{1}{k - 1}\sum_{i}^ {k - 1} \mathrm{sim}(d_i, Q) - \mathrm{sim}(d_{i+1}, Q)
\end{equation}

The adaptive weighting parameter $\alpha$ is then dynamically determined based on $P$ using a Sigmoid function:
\begin{equation}
\alpha = 1 - \sigma(P) = 1- \frac{1}{1 + e^{-P}}
\end{equation}
The Sigmoid function, $\sigma(\cdot)$, ensures that $\alpha$ remains bounded within the range $[0, 1]$, controlling the weight of the retriever during the joint optimization.

This dynamic adjustment mechanism empowers Joint-GCG to adaptively balance the optimization of retrieval and generation objectives, resulting in more potent and robust poisoning attacks across diverse situations.

\subsubsection{Step 2: Candidate Set Generation And Re-evaluation}
Using the joint gradient $\nabla L_\text{joint}$, we identify the most promising token substitutions. For each position in the adversarial sequence $S_\text{adv}$, we search for the Top-$N$ tokens from the vocabulary that would most likely decrease $L_\text{joint}$. Then we randomly select $M$ positions to perform the replacement and generate a set of candidate sequences $S'_\text{adv}$. For each set in the set, we perform a forward pass involving recalculating both $L_\text{gen}$ and $L_\text{ret}$, then calculate $L_\text{joint}$ for all candidates using Equation~\ref{eq:joint_loss}.

\subsubsection{Step 3: Greedy Selection and Update}
Finally, we perform a greedy selection. From all evaluated candidates, we choose the one that yields the lowest $L_\text{joint}$ and update $S_\text{adv}$ accordingly for the next iteration.
\begin{equation}
S_{\text{adv}}^{(t+1)} = \underset{S'_{\text{adv}} \in \text{Candidates}}{\arg\min} L_{\text{joint}}(S'_{\text{adv}})
\end{equation}
After multiple optimization iterations, we can achieve the Joint Optimization Objective in Eq~\ref{eq:obj}, and simultaneously ensure the poisoned document is retrieved and the generator's output is manipulated.

\section{Experiments}
\subsection{Experimental Setup}
Joint-GCG's effectiveness was rigorously assessed against state-of-the-art baselines (PoisonedRAG with GCG~\cite{zou2023universal}, LIAR~\cite{tan2024glue}, Phantom~\cite{chaudhari2024phantom}) across diverse conditions, including targeted and batch (see Appendix~\ref{exp: 4}) query poisoning, ablation studies, synthetic corpora, and black-box transferability.

\subsubsection{Datasets And Models} Evaluations utilized three open-domain Question-Answering (QA) datasets: \textbf{MS MARCO}~\cite{nguyen2016ms}, \textbf{Natural Questions (NQ)}~\cite{kwiatkowski2019natural}, and \textbf{HotpotQA}~\cite{yang2018hotpotqa}. The synthetic corpus (GPT-4o-mini generated) simulated retrieval for AWF calculations for the Synthetic Corpus experiment. Two dense retrieval models, \textbf{Contriever}~\cite{izacard2021unsupervised} and \textbf{BGE}~\cite{xiao2024c}, along with two LLM generators, \textbf{Llama3-8B}~\cite{dubey2024llama} and \textbf{Qwen2-7B}~\cite{yang2024qwen2technicalreport}, were used.

\subsubsection{Metrics}
We use the following metrics to evaluate the effectiveness of the poisoning attacks:
\begin{itemize}
    \item \textbf{Retrieval Attack Success Rate ($ASR_\text{ret}$):} The percentage of target queries for which the poisoned document is retrieved within the top-$k$ results.
    \item \textbf{Generation Attack Success Rate ($ASR_\text{gen}$):} The percentage of target queries for which the LLM generates the desired target output, i.e., the generated output contains the target. We use this approach to align with PoisonedRAG~\cite{zou2024poisonedrag}, as it shows negligible differences from human evaluation.
    \item \textbf{Position of Poisoned Document ($Pos_{p}$):} The average rank ($1 \le Pos_p \le k$) of the poisoned document in the retrieval results for the target queries. Lower values indicate a stronger positioning of the poisoned document.
\end{itemize}

Experiments were repeated three times. In experiments comparing Joint-GCG to PoisonedRAG with GCG and LIAR, we set the length of $S_\text{adv}$ to 32 tokens and the optimization iterations to 64. We employed a variant of GCG, MCG~\cite{chaudhari2024phantom}, to enhance the attack efficiency and utilized its default hyperparameter. We expressly set the candidate numbers of each step to 128 and token positions $N$ to 16, used ASCII-character-only tokens for attacks, and configured the optimization target to the incorrect answer. For the LIAR method, we used a 1:1 ratio for the $S_\text{adv}$ length (16 tokens each for the retriever and generator) and optimization iterations (8 steps each for the retriever and generator), attacking the retriever first, followed by the generator.

\begin{table*}[htb]
\caption{$ASR$ and mean $pos_p$ of GCG, LIAR, and Joint-GCG at 64 optimization steps. Parenthetical values show $ASR_\text{gen}$ on queries where unoptimized attacks failed.}
\centering
\resizebox{0.9 \linewidth}{!}{
\begin{tabular}{@{}ccc|cc|cc|cc@{}}
\toprule
\multirow{2}{*}{\textbf{Retriever}} & \multirow{2}{*}{\textbf{Metrics}} & \textbf{Dataset} & \multicolumn{2}{c|}{\textbf{MS MARCO}} & \multicolumn{2}{c|}{\textbf{NQ}} & \multicolumn{2}{c}{\textbf{HotpotQA}} \\ \cmidrule(l){3-9} 
 &  & \textbf{Attack / LLM} & Llama3 & Qwen2 & Llama3 & Qwen2 & Llama3 & Qwen2 \\ \midrule
\multirow{10}{*}{Contriever} & \multicolumn{1}{c|}{\multirow{3}{*}{$ASR_\text{ret}$}} & GCG & 96.00\% & 95.67\% & 72.00\% & 72.00\% & 94.33\% & 97.00\% \\
 & \multicolumn{1}{c|}{} & LIAR & \textbf{100.00\%} & \textbf{100.00\%} & 93.33\% & 96.33\% & 99.00\% & \textbf{100.00\%} \\
 & \multicolumn{1}{c|}{} & Joint-GCG & \textbf{100.00\%} & \textbf{100.00\%} & \textbf{99.00\%} & \textbf{99.00\%} & \textbf{100.00\%} & \textbf{100.00\%} \\ \cmidrule(l){2-9} 
 & \multicolumn{1}{c|}{\multirow{4}{*}{$ASR_\text{gen}$}} & GCG & 90.0\% (76.7\%) & 91.0\% (80.0\%) & 72.0\% (41.5\%) & 70.0\% (39.0\%) & 90.3\% (76.7\%) & 97.0\% (87.5\%) \\
 & \multicolumn{1}{c|}{} & LIAR & 89.0\% (74.4\%) & 95.3\% (88.9\%) & 89.0\% (73.2\%) & 86.0\% (68.3\%) & 92.0\% (81.4\%) & 98.0\% (91.7\%) \\
 & \multicolumn{1}{c|}{} & Joint-GCG & \textbf{94.0\% (86.0\%)} & \textbf{96.3\% (91.1\%)} & \textbf{92.0\% (82.9\%)} & \textbf{95.0\% (87.8\%)} & \textbf{97.3\% (93.0\%)} & \textbf{99.0\% (95.8\%)} \\
 & \multicolumn{1}{c|}{} & w/o optimize & 51.0\% & 49.0\% & 50.0\% & 34.0\% & 59.0\% & 60.0\% \\ \cmidrule(l){2-9} 
 & \multicolumn{1}{c|}{\multirow{3}{*}{$Pos_p\downarrow$}} & GCG & 1.36 & 1.43 & 2.59 & 2.56 & 1.46 & 1.2 \\
 & \multicolumn{1}{c|}{} & LIAR & 1.13 & 1.08 & 1.52 & 1.43 & 1.14 & 1.06 \\
 & \multicolumn{1}{c|}{} & Joint-GCG & \textbf{1.01} & \textbf{1.05} & \textbf{1.25} & \textbf{1.22} & \textbf{1.04} & \textbf{1.01} \\ \midrule
\multirow{10}{*}{BGE} & \multicolumn{1}{c|}{\multirow{3}{*}{$ASR_\text{ret}$}} & GCG & 74.00\% & 73.30\% & 96.00\% & 98.67\% & 100.00\% & 100.00\% \\
 & \multicolumn{1}{c|}{} & LIAR & \textbf{99.00\%} & 97.30\% & \textbf{100.00\%} & \textbf{100.00\%} & 100.00\% & 100.00\% \\
 & \multicolumn{1}{c|}{} & Joint-GCG & \textbf{99.00\%} & \textbf{99.00\%} & \textbf{100.00\%} & \textbf{100.00\%} & 100.00\% & 100.00\% \\ \cmidrule(l){2-9} 
 & \multicolumn{1}{c|}{\multirow{4}{*}{$ASR_\text{gen}$}} & GCG & 68.0\% (60.7\%) & 67.0\% (57.1\%) & \textbf{93.0\% (89.1\%)} & 97.0\% (95.5\%) & 98.0\% (95.9\%) & \textbf{99.0\% (97.4\%)} \\
 & \multicolumn{1}{c|}{} & LIAR & 83.7\% (78.6\%) & \textbf{92.0\% (85.7\%)} & 89.3\% (80.0\%) & 93.0\% (86.4\%) & 93.7\% (85.7\%) & 96.0\% (89.5\%) \\
 & \multicolumn{1}{c|}{} & Joint-GCG & \textbf{87.0\% (85.7\%)} & \textbf{92.0\% (85.7\%)} & \textbf{93.0\% (87.3\%)} & \textbf{97.7\% (95.5\%)} & \textbf{99.0\% (98.0\%)} & \textbf{99.0\% (97.4\%)} \\
 & \multicolumn{1}{c|}{} & w/o optimize & 31.0\% & 27.0\% & 39.0\% & 31.0\% & 46.0\% & 46.0\% \\ \cmidrule(l){2-9} 
 & \multicolumn{1}{c|}{\multirow{3}{*}{$Pos_p\downarrow$}} & GCG & 2.87 & 3.02 & 1.36 & 1.23 & 1.04 & 1.01 \\
 & \multicolumn{1}{c|}{} & LIAR & 1.5 & 1.69 & \textbf{1.04} & \textbf{1.07} & \textbf{1.01} & 1.01 \\
 & \multicolumn{1}{c|}{} & Joint-GCG & \textbf{1.38} & \textbf{1.47} & 1.06 & \textbf{1.07} & \textbf{1.01} & 1.01 \\ \bottomrule
\end{tabular}%
    }
\label{tab: target-query-64}
\end{table*}

\subsection{Experiment Results}

\subsubsection{Targeted Query Poisoning: Baseline Comparison}
\label{exp: 0}
\begin{figure}[t]
  \centering
  \includegraphics[width=0.75\columnwidth]{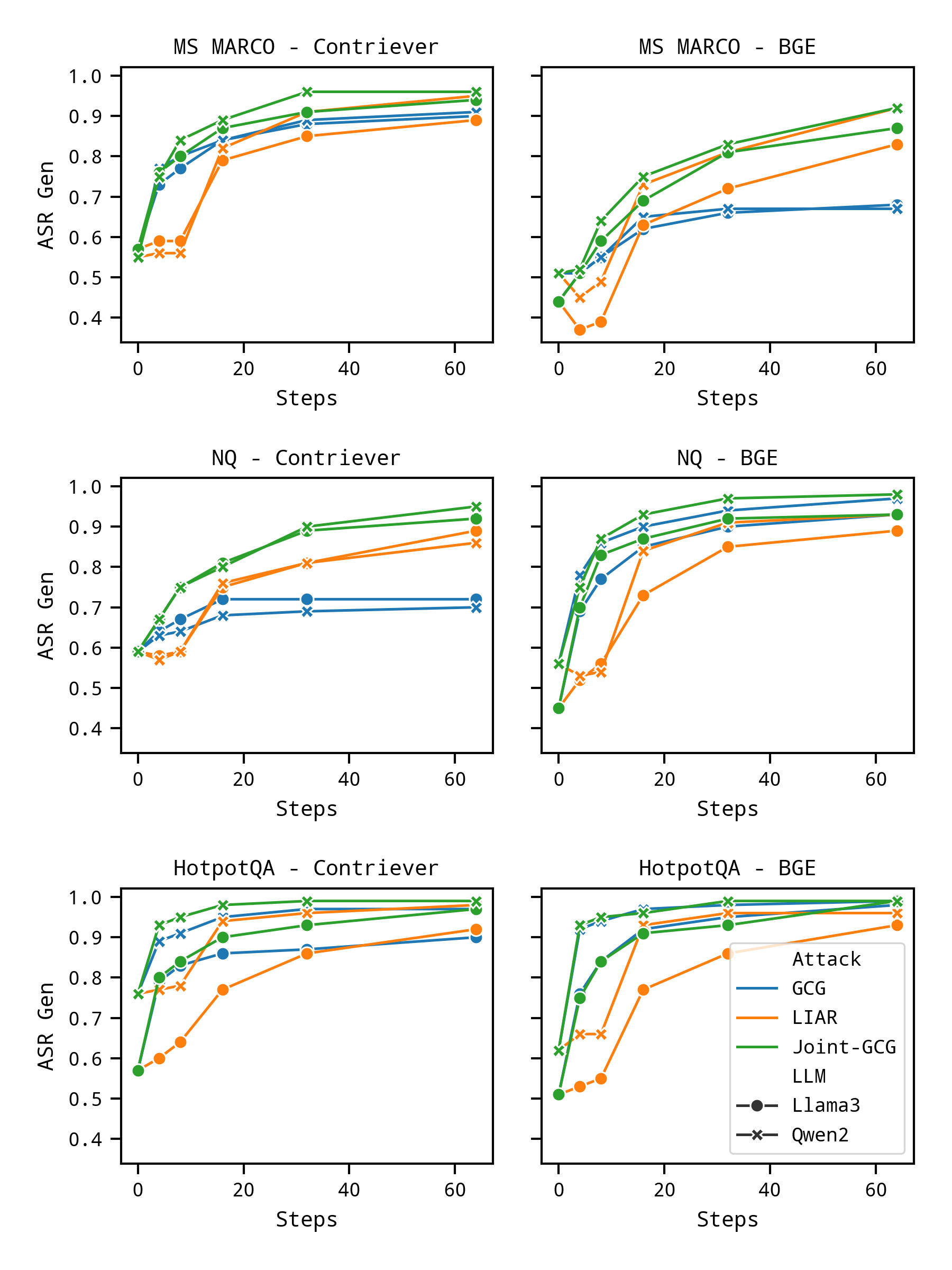}
  \caption {$ASR_\text{gen}$ of GCG, LIAR, and Joint-GCG at various optimization steps.}
  \label{fig: target-query-asr-vs-ksteps}
\end{figure}

Joint-GCG consistently outperforms GCG and LIAR in targeted query poisoning, achieving higher $ASR_\text{ret}$ and $ASR_\text{gen}$ across diverse settings (Table~\ref{tab: target-query-64}). Joint-GCG maintains near-perfect $ASR_\text{ret}$ ($100\%$) for Llama3 and Qwen2 across all datasets, significantly surpassing baselines in generation attacks, particularly on NQ and HotpotQA. For instance, Joint-GCG yields up to $99.0\%$ $ASR_\text{gen}$ for Llama3 and $95.8\%$ for Qwen2 on HotpotQA, outperforming GCG and LIAR by several percentage points. To isolate the impact of the optimization process itself, we also calculated the ASR on the subset of queries where a simple, unoptimized poison initially failed. The results shown in parentheses in Table~\ref{tab: target-query-64} highlight the significant gains achieved by our method. Figure~\ref{fig: target-query-asr-vs-ksteps} further illustrates Joint-GCG's superior efficacy, achieving comparable or higher $ASR_\text{gen}$ with significantly fewer optimization steps, underscoring its enhanced efficiency and effectiveness. This performance gain stems from Joint-GCG's integrated approach, preventing retrieval degradation and ensuring consistent poisoning efficacy.

Also, Joint-GCG has better efficacy (i.e., achieves higher $ASR_\text{gen}$ at fewer optimization steps). Figure~\ref{fig: target-query-asr-vs-ksteps} visually confirms Joint-GCG's superior efficacy, demonstrating that it reaches comparable or higher $ASR_\text{gen}$ than GCG and LIAR while requiring significantly fewer optimization steps. This faster convergence underscores Joint-GCG's enhanced efficiency and effectiveness in targeted query poisoning attacks.

Joint-GCG's performance gains stem from its innovative approach of integrating the retriever and generator gradients. This effectively prevents retrieval degradation and ensures the efficacy and success of poisoning throughout the optimization process. The consistently higher index ranking of the poisoned corpus highlights Joint-GCG's strengths as a leading method for targeted query poisoning, especially in adversarial contexts where high document rankings are essential for effective generator manipulation.

\subsubsection{Ablation Study}
\label{exp: 1}
\begin{figure}[t]
  \centering
  \includegraphics[width=0.8\columnwidth]{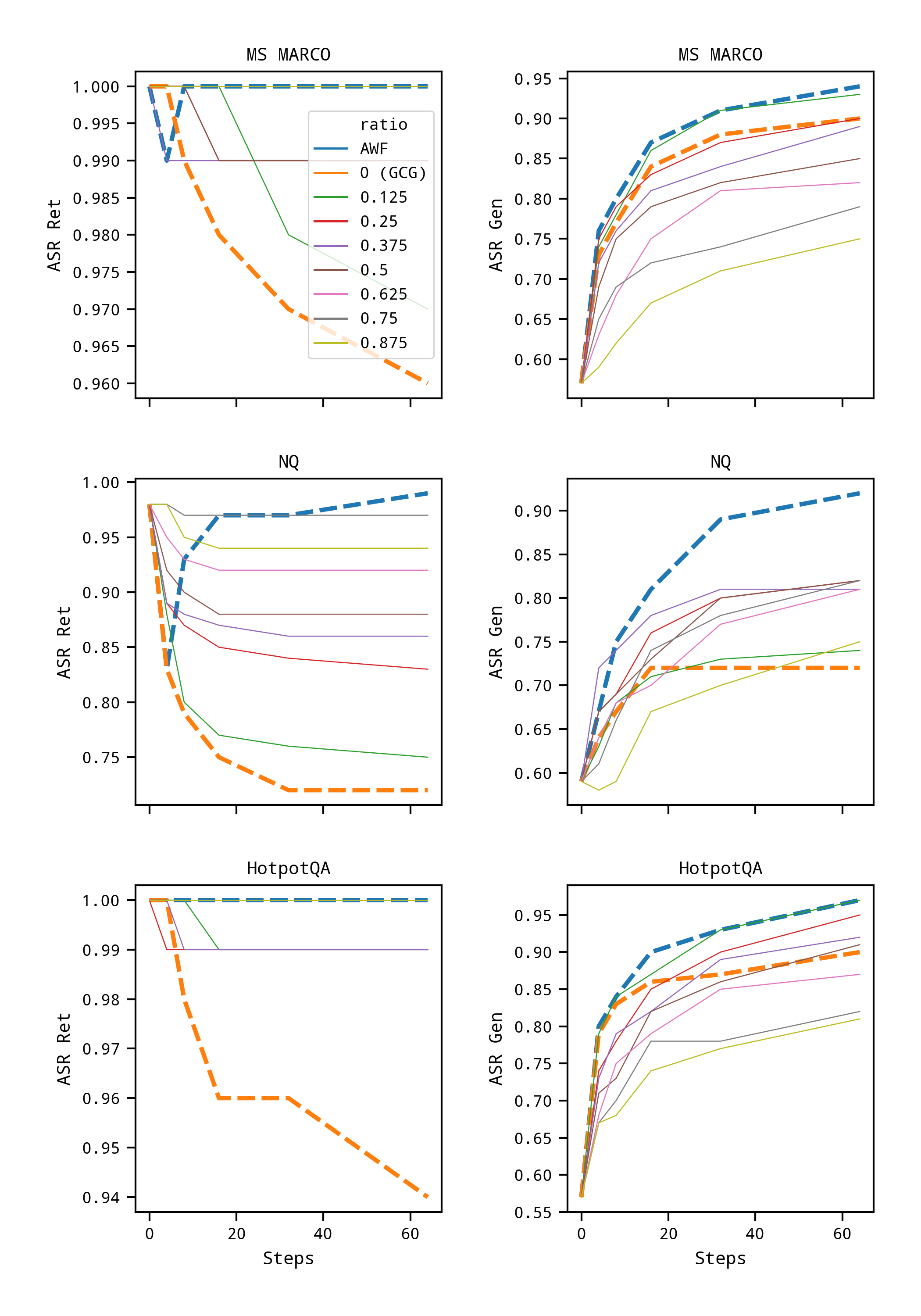}
  \caption {$ASR$ of Joint-GCG with various fixed weights and AWF, using Contriver as the retriever and Llama3 as the generator.}
  \label{fig: awf-ablation}
\end{figure}

An ablation study analyzed the contribution of Joint-GCG's core components (Table~\ref{tab: ablation}). Removing Cross-Vocabulary Projection (CVP) and Gradient Tokenization Alignment (GTA) resulted in a modest but consistent 2\% average decrease in $ASR_\text{gen}$, indicating their role in enhancing attack potency by enabling proper gradient fusion. More significantly, the removal of the retriever-side loss ($L_\text{ret}$) led to a pronounced decrease in $ASR_\text{gen}$ across all datasets and generators (e.g., a 3-5\% drop on MS MARCO), underscoring its crucial role in guiding optimization towards potent poisoned documents. Furthermore, Adaptive Weighted Fusion (AWF) consistently demonstrated superior or comparable performance to all fixed retrieval-generation gradient weights (Figure~\ref{fig: awf-ablation}), highlighting its effectiveness in dynamically balancing retriever and generator optimization.
\begin{table}[ht]
\centering
\caption{$ASR_\text{gen}$ with CVP and GTA removed and $ASR_\text{gen}$ with $L_\text{gen}$ only across datasets and generators, using Contriever as the retriever.}
\label{tab: ablation}
\resizebox{0.8\columnwidth}{!}{%
\begin{tabular}{@{}cccc@{}}
\toprule
\textbf{Dataset} & \textbf{Settings} & \textbf{Llama3} & \textbf{Qwen2} \\ \midrule
\multirow{4}{*}{MS MARCO} & Full Joint-GCG & \textbf{94.00\%} & \textbf{96.33\%} \\
 & w/o CVP + GTA & 93.33\% & 96.00\% \\
 & w/o $L_\text{ret}$ & 91.00\% & 92.33\% \\
 & Base (GCG) & 90.00\% & 91.00\% \\ \midrule
\multirow{4}{*}{NQ} & Full Joint-GCG & \textbf{92.00\%} & \textbf{95.00\%} \\
 & w/o CVP + GTA & 91.00\% & 93.00\% \\
 & w/o $L_\text{ret}$ & 86.67\% & 94.00\% \\
 & Base (GCG) & 72.00\% & 70.00\% \\ \midrule
\multirow{4}{*}{HotpotQA} & Full Joint-GCG & \textbf{97.33\%} & \textbf{99.00\%} \\
 & w/o CVP + GTA & 95.00\% & \textbf{99.00\%} \\
 & w/o $L_\text{ret}$ & 91.33\% & 98.67\% \\
 & Base (GCG) & 90.00\% & 97.00\% \\ \bottomrule
\end{tabular}%
}
\end{table}

\begin{table}[ht]
\caption{$ASR$s on MS MARCO dataset with real top-k retrieval and synthetic corpus-based AWF.}
\centering
\resizebox{0.75\columnwidth}{!}{
\begin{tabular}{@{}ccccc@{}}
\toprule
\textbf{Retriever} & \textbf{Metrics} & \textbf{Settings} & \textbf{Llama3} & \textbf{Qwen2} \\ \midrule
\multirow{6}{*}{Contriever} & \multirow{3}{*}{$ASR_\text{ret}$} & Real & 100.00\% & 100.00\% \\
 &  & Synthetic & 100.00\% & 100.00\% \\
 &  & w/o optimize & 98.00\% & 98.00\% \\ \cmidrule(l){2-5} 
 & \multirow{3}{*}{$ASR_\text{gen}$} & Real & 94.00\% & 96.33\% \\
 &  & Synthetic & 62.00\% & 58.00\% \\
 &  & w/o optimize & 51.00\% & 49.00\% \\ \midrule
\multirow{6}{*}{BGE} & \multirow{3}{*}{$ASR_\text{ret}$} & Real & 99.00\% & 99.00\% \\
 &  & Synthetic & 89.33\% & 84.00\% \\
 &  & w/o optimize & 70.00\% & 70.00\% \\ \cmidrule(l){2-5} 
 & \multirow{3}{*}{$ASR_\text{gen}$} & Real & 87.00\% & 92.00\% \\
 &  & Synthetic & 41.00\% & 36.67\% \\
 &  & w/o optimize & 31.00\% & 27.00\% \\ \bottomrule
\end{tabular}%
}
\label{tab: synthetic-corpus-experiment}
\end{table}
\subsubsection{Synthetic Corpus: Removing Top-k Retrieval Dependency}
\label{exp: 2}
Joint-GCG's practicality was assessed by using a synthetic corpus to simulate retrieval, eliminating dependency on real-world top-$k$ results (Table~\ref{tab: synthetic-corpus-experiment}). While some performance variance exists compared to real corpus data, $ASR_\text{gen}$ using synthetic data remains commendable (e.g., 62\% for Contriever/Llama3), demonstrating a solid attack capability in restricted settings.

\subsubsection{Evaluating Poison Generalization}
\label{exp: 3}

\begin{figure}[h]
    \centering
    \includegraphics[width=0.95\columnwidth]{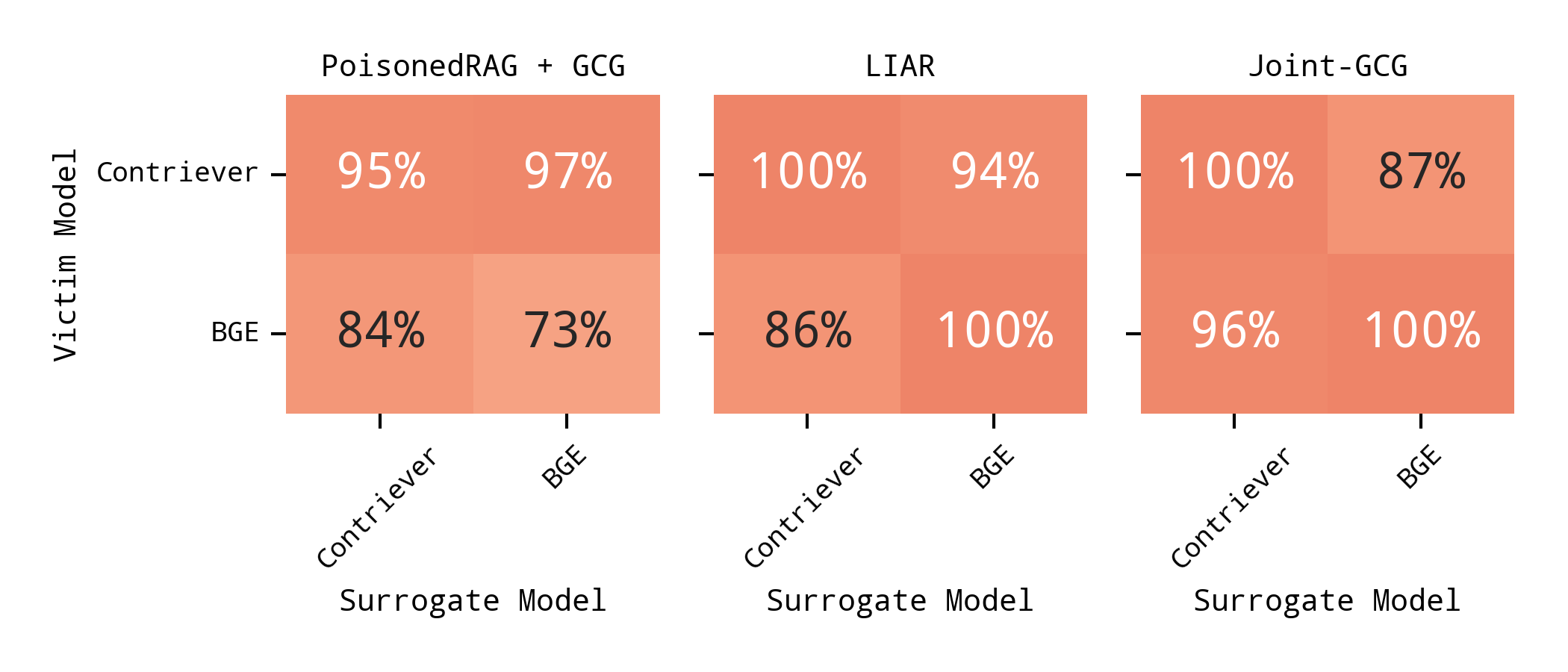}
    \caption{$ASR_\text{ret}$ on various victim retrievers with Llama3 generator, on MS MARCO.}
    \label{fig: ret_transferability_heatmap}
\end{figure}

\begin{figure}[h]
    \centering
    \includegraphics[width=0.95\columnwidth]{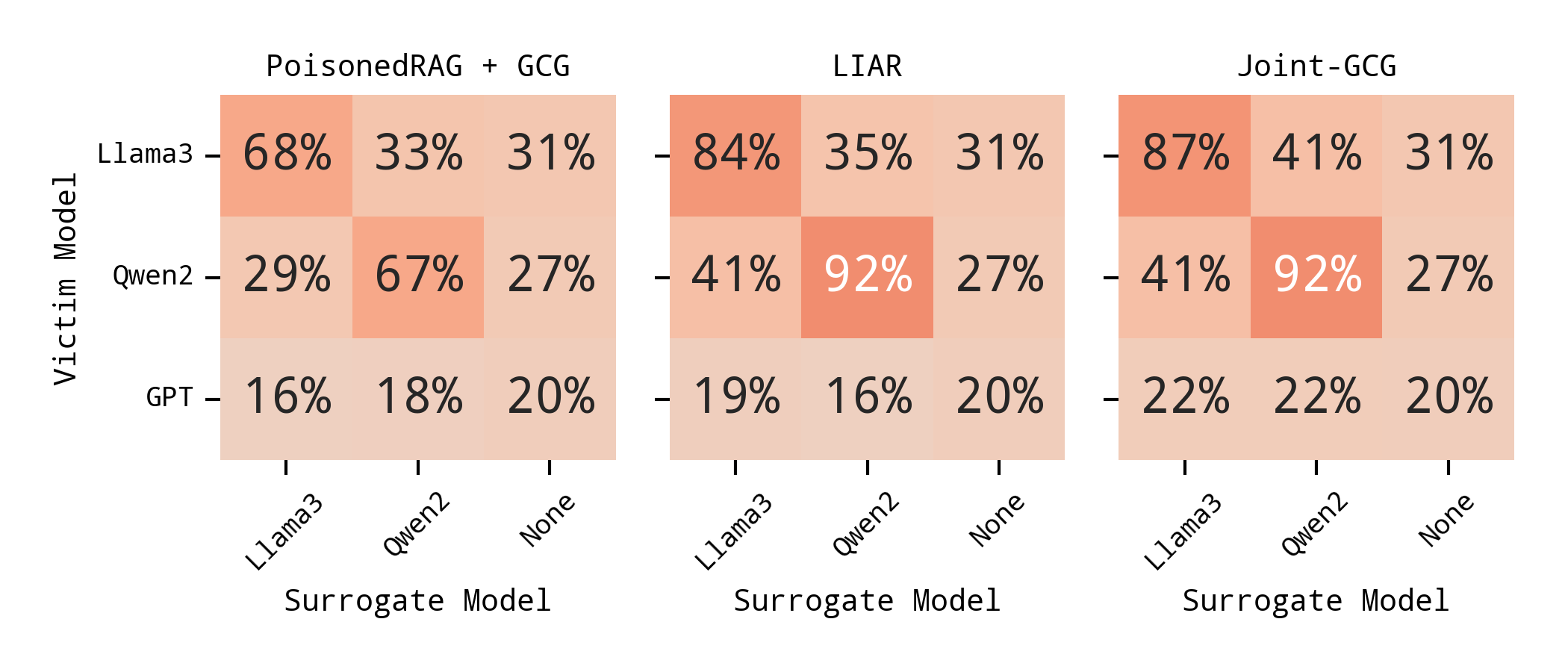}
    \caption{$ASR_\text{gen}$ on various victim generators, `None’ as surrogate model represents the unoptimized initial samples.}
    \label{fig: gen_transferability_heatmap}
\end{figure}

We evaluated the generalization of Joint-GCG's poisons, which is critical for assessing real-world threats.
\paragraph{\textbf{Cross-Generator Transferability}} Joint-GCG also exhibited consistent cross-generator transferability (Figure~\ref{fig: gen_transferability_heatmap}) with the BGE retriever, on the MS MARCO dataset. Poisons optimized for Llama3 achieved 41\% $ASR_\text{gen}$ on Qwen2, and vice versa. While LIAR showed similar transfer from Qwen2 to Llama3 (41\%), it was slightly less from Llama3 to Qwen2 (35\%). Notably, Joint-GCG's optimization-driven transferability extended to black-box commercial LLMs like GPT-4o, showing a noticeable 2\% increase in attack success compared to non-optimized poisons, a phenomenon unseen in prior methods. These findings highlight that RAG systems face a broader, more generalized attack surface, as attackers can enhance cross-generator poison transferability through targeted optimization on readily available surrogate models.
\paragraph{\textbf{Cross-Retriever Transferability}} As shown in Figure~\ref{fig: ret_transferability_heatmap}, poisons optimized by Joint-GCG demonstrated high cross-retriever transferability, achieving 96\% $ASR_\text{ret}$ from BGE to Contriever and 87\% $ASR_\text{ret}$ from Contriever to BGE. TThis strong, bidirectional performance is highly competitive with LIAR (which achieved 94\% and 86\% in the same respective transfer scenarios) and PoisonedRAG + GCG (97\% and 84\%). Unlike the PoisonedRAG + GCG baseline which failed to achieve perfect retrieval (showing 95\% and 73\% $ASR_\text{ret}$), Joint-GCG attained a 100\% $ASR_\text{ret}$ in its direct attacks, matching LIAR. This combination of perfect direct efficacy and strong transferability underscores the robustness and practical threat of Joint-GCG's poisons.

\section{Conclusion}

We present \textbf{Joint-GCG}, a framework that elevates RAG poisoning via unified retrieval-generation gradient-based optimization. By harmonizing retrieval and generation objectives through Cross-Vocabulary Projection, Gradient Tokenization Alignment, and Adaptive Weighted Fusion, Joint-GCG overcomes the disjointed nature of prior attacks. Our framework consistently outperformed prior art, delivering substantial gains in attack success rates—up to 25\% in certain scenarios—and demonstrating superior efficiency. Ablations confirm the role of each component, while synthetic corpus tests and poison generalization experiments demonstrate the broad applicability. The framework's potency in batch poisoning further underscores its practical threat. Joint-GCG provides a robust framework for understanding and mitigating the evolving threat landscape of RAG-based applications. 

\section*{Ethical Statement}
The Joint-GCG framework, while advancing the understanding of RAG system vulnerabilities to improve security, presents potential misuse risks. We have prioritized responsible disclosure, striking a balance between scientific transparency and these concerns. Our research, conducted on controlled datasets and non-production systems, aims to proactively identify vulnerabilities, motivating the development of robust defenses and encouraging the design of security-first RAG systems. We strongly advocate for using these findings for security research and system improvement, not exploitation. Organizations deploying RAG systems should implement comprehensive security measures, including regular audits, content filtering, and continuous monitoring, as understanding these vulnerabilities is crucial to building more secure AI applications.

\section*{Acknowledgments}
This work was supported by the National Key Research and Development Program of China (No. 2024YFF0618800), the National Natural Science Foundation of China Grant  No. 62402484, the Basic Research Program of ISCAS Grant No. ISCAS-JCZD-202405, the China Southern Power Grid Company Limited ZBKJXM20240173, and the Youth Innovation Promotion Association Chinese Academy of Sciences. The authors wish to thank Mr. Zhu Qiming for his valuable insights during the conceptualization of the joint-attack methods. They are also grateful to the anonymous reviewers for their constructive feedback.


\bibliography{cite} 


\appendix

\section{Related Works}
\label{A-related-works}
\subsection{Retrieval-Augmented Generation (RAG) Systems}
Retrieval-Augmented Generation (RAG) systems represent a significant advancement in mitigating the inherent limitations of Large Language Models (LLMs), such as knowledge cutoffs and hallucinations ~\cite{lewis2020retrieval, gao2024retrievalaugmentedgenerationlargelanguage}. The core principle of RAG is to ground LLM responses in external, verifiable knowledge. A typical RAG architecture involves a retriever and a generator. When a query is received, the retriever first fetches relevant documents or data snippets from a large external corpus (e.g., a vector database, enterprise knowledge base, or the internet)~\cite{asai2023retrieval}. These retrieved contexts are then provided to the LLM (the generator) along with the original query. The LLM synthesizes this information to produce a more accurate, timely, and contextually appropriate response ~\cite{gao2024retrievalaugmentedgenerationlargelanguage}. This reliance on external corpora, while beneficial for accuracy and currency, introduces new attack surfaces, particularly corpus poisoning, which is the focus of our work.

\subsection{Adversarial Attacks on Large Language Models}
Large Language Models, despite their impressive capabilities, are susceptible to various adversarial attacks ~\cite{weidinger2021ethical}. These vulnerabilities include prompt injection, where malicious instructions are embedded in the input to hijack the model's output~\cite{greshake2023more}, and training data poisoning of the base LLM itself, which can introduce subtle biases or backdoors~\cite{chakraborty2018adversarial}. These general vulnerabilities underscore the necessity for robust security measures throughout all stages of LLM development and deployment.

\subsubsection{Gradient-Based Optimization for Adversarial Text Generation}

Gradient-based optimization has become a cornerstone for crafting adversarial examples against LLMs. Early methods, such as \textit{HotFlip}~\cite{ebrahimi2017hotflip}, utilized gradients related to input tokens to identify minimal character-level or token-level perturbations that could alter model predictions. Simple substitution techniques~\cite{li2018textbugger} and heuristic optimizations~\cite{li2020bert} have achieved only modest success, primarily with smaller models. In contrast, gradient-guided approaches have shown superior efficacy against robust transformer architectures. The \textit{Greedy Coordinate Gradient (GCG)} attack~\cite{zou2023universal} extended this by optimizing a universal adversarial suffix to elicit desired (often harmful) responses. \textit{Multi Coordinate Gradient (MCG)}, proposed in the Phantom work~\cite{chaudhari2024phantom}, further refines this by considering multiple substitutions simultaneously for efficiency. Improved versions, such as I-GCG~\cite{jia2024improved}, have also been explored. These methods demonstrate the power of gradient information but are primarily designed for direct attacks on the LLM, not the complex two-stage RAG pipeline.

\subsection{Data Poisoning Attacks}

\subsubsection{Classical Data Poisoning in Machine Learning}
Data poisoning is a type of malicious attack in which an attacker manipulates the training data of a machine learning model to compromise its behavior during inference~\cite{goldblum2022dataset}. By injecting a small quantity of carefully crafted malicious samples into the training dataset, attackers primarily seek to degrade overall model performance, cause misclassification for targeted inputs, or implant backdoors triggered by specific inputs~\cite{shumailov2021manipulating, liu2019data}.

\subsubsection{Attacks on Information Retrieval Systems}
Information Retrieval (IR) systems have also been targets of manipulation. Traditional search engines faced "spamdexing" or "search engine poisoning," where malicious actors used techniques like keyword stuffing, hidden text, and link farms to artificially boost the ranking of certain web pages~\cite{gyongyi2005web, bevendorff2024google}. Modern retrieval components, especially dense retrievers used in RAG, can be misled by imperceptible perturbations to documents, leading to ranking manipulation~\cite{farooq2019survey, mallen2022not}. Malicious attacks can force rankers to incorrectly order documents or retrieve irrelevant content, even with minimal modifications to the corpus.

\subsubsection{Corpus Poisoning in RAG Systems}
The reliance of RAG systems on external corpora makes them uniquely vulnerable to corpus poisoning. Existing research has explored various strategies: \textit{PoisonedRAG}~\cite{zou2024poisonedrag} focused on optimizing poisoned documents to maximize their retrieval probability. \textit{HijackRAG}~\cite{zhang2024hijackrag} applied similar principles to prompt leaking and spam generation. However, optimizing solely for retrieval can compromise the linguistic qualities needed to effectively steer the generator, often necessitating a larger number of injected documents.
Sequential optimization approaches, such as applying \textit{HotFlip} for retrieval and then \textit{GCG} for generation, can be suboptimal because modifications made in one stage may negatively impact the other. \textsc{LIAR}~\cite{tan2024glue} attempted to achieve better integration through an iterative loop but still lacks proper joint optimization, as it pre-assigns fixed optimizable lengths and optimization steps. Phantom \cite{chaudhari2024phantom} introduced trigger-based batch poisoning, but optimized retrieval and generation sequentially, which sometimes led to retrieval failures. \textit{TrojanRAG}~\cite{cheng2024trojanrag} targets the retriever model itself, a different threat model requiring greater access.
In contrast, our Joint-GCG framework addresses these limitations by performing a truly joint optimization across both the retriever and the generator, marking a departure from these disjointed or sequential strategies.

\section{Detailed Methodology}
This section expands on the technical details of the Joint-GCG framework, providing deeper insights into its core innovations.

\subsection{Cross-Vocabulary Projection (CVP) Details}
\label{A-CVP-additional}

\subsubsection{Motivation \& Overview}
Cross-Vocabulary Projection (CVP) addresses the fundamental vocabulary mismatch between retrievers and generators in RAG systems. Since retrievers and generators are typically pre-trained independently, their tokenization schemes, vocabularies, and embedding spaces differ significantly. This discrepancy prevents direct gradient alignment between the two components. CVP bridges this gap by learning a joint embedding space through an autoencoder trained on shared tokens and deriving a linear transformation matrix to project retriever gradients into the generator's embedding space.

\subsubsection{Autoencoder Architecture \& Training}
The CVP autoencoder consists of an encoder-decoder pair with multiple ReLU-activated dense layers (Figure~\ref{fig:cvp-arch}):

\begin{figure}[h]
    \centering
    \includegraphics[width=0.8\columnwidth]{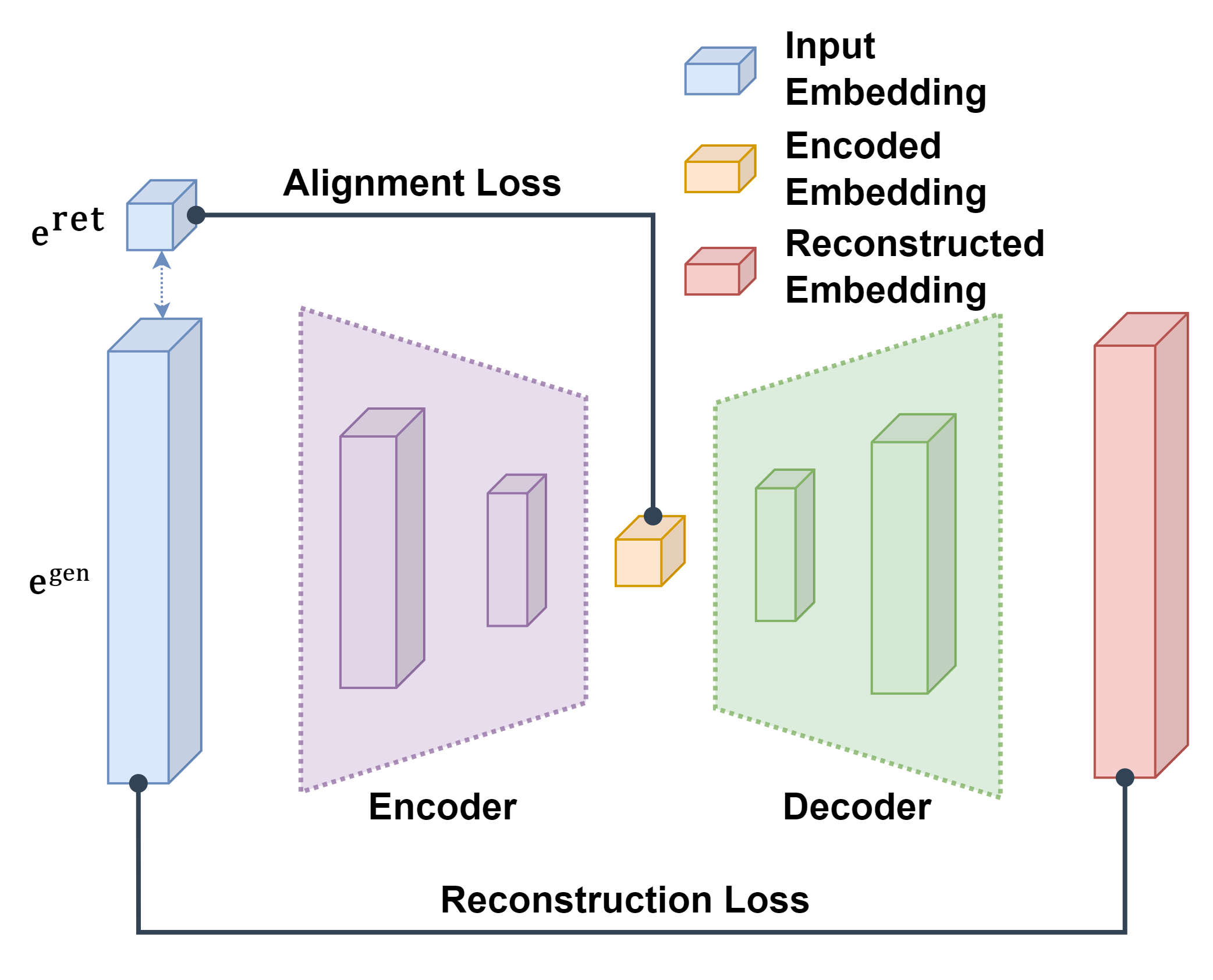}
    \caption{CVP autoencoder architecture. Token embeddings from the generator (LLM) are encoded into the retriever's space and decoded back.}
    \label{fig:cvp-arch}
\end{figure}

\textbf{Encoder:} Maps the generator's embeddings ($\mathbb{R}^{D_\text{gen}}$) to the retriever's embedding space ($\mathbb{R}^{D_\text{ret}}$) via:
\begin{align}
    h_1 &= \text{ReLU}(W_0 x + b_0) \\
    h_2 &= \text{ReLU}(W_1 h_1 + b_1) \\
    \text{Enc}(e_\text{gen}) &= W_2 h_2 + b_2
\end{align}
where $e_\text{gen}$ is a generator token embedding, $W_0 \in \mathbb{R}^{2048 \times D_\text{gen}}$, $W_1 \in \mathbb{R}^{1024 \times 2048}$, and $W_2 \in \mathbb{R}^{D_\text{ret} \times 1024}$.

\textbf{Decoder:} Reconstructs the generator's embeddings from encoded retriever-space embeddings:
\begin{align}
    h'_1 &= \text{ReLU}(W'_0 y + b'_0) \\
    h'_2 &= \text{ReLU}(W'_1 h'_1 + b'_1) \\
    \text{Dec}(e_\text{ret}) &= W'_2 h'_2 + b'_2
\end{align}
where $e_\text{ret}$ is the corresponding retriever token embedding, $W'_0 \in \mathbb{R}^{1024 \times D_\text{ret}}$, $W'_1 \in \mathbb{R}^{2048 \times 1024}$, and $W'_2 \in \mathbb{R}^{D_\text{gen} \times 2048}$.

We collect the shared tokens from both models and create a train-test split with an 80\% training set and a 20\% validation set.
The model is trained on shared token pairs $(e^\text{gen}, e^\text{ret})$ in the training set using a composite loss:
\begin{equation}
    \mathcal{L} = \alpha \mathcal{L}_{\text{rec}} + (1-\alpha) \mathcal{L}_{\text{align}}
\end{equation}
where:
\begin{align}
    \mathcal{L}_{\text{rec}} &= \sum_{i} \| \text{Dec}(\text{Enc}(e^\text{gen}_i)) - e^\text{gen}_i \|_2 \\
    \mathcal{L}_{\text{align}} &= \sum_{i} \| \text{Enc}(e^\text{gen}_i) - e^\text{ret}_i \|_2
\end{align}
with $\alpha=0.25$. We use AdamW with cosine annealing (initial learning rate $10^{-5}$) to train the model for 500 epochs at most, with an early stopping on validation loss.

\subsubsection{Transfer Matrix via Least Squares}
After autoencoder training, we compute a linear projection matrix $W \in \mathbb{R}^{V_\text{ret} \times V_\text{gen}}$ to map retriever token gradients into the generator's gradient space as below:

\begin{enumerate}
    \item \textbf{Encode Generator Embeddings:} We first compute the encoded embedding $\tilde E_\text{gen} \in \mathbb{R}^{V_\text{gen} \times D_\text{ret}}$ for the generator embedding $E_\text{gen} \in \mathbb{R}^{V_\text{gen} \times D_\text{gen}}$ to match embedding dimensions $D_\text{gen}$ and $D_\text{ret}$:
    \begin{equation}
        \tilde{E}_\text{gen} = \text{Enc}(E_\text{gen})
    \end{equation}

    \item \textbf{Solve Least Squares:} We employ the least squares method to find the optimal projection matrix between tokens. Specifically, for each retriever token embedding $y$, we find weights $W_i$ over generator tokens that minimize:
    \begin{equation}
        \operatorname*{argmin}_{W_i} \| W_i \tilde{E}_\text{gen} - y \|_2^2
    \end{equation} 
\end{enumerate}
We solve the $W_i$ via PyTorch's \texttt{torch.lstsq}.

Finally, concatenating $W_i$ for each $y$ yields a projection matrix $W$ that maps retriever token influence into the generator's embedding space, enabling joint gradient optimization. 

\subsubsection{Evaluation}
To evaluate the effectiveness of the CVP autoencoder, we conducted a series of experiments on the validation set. We assessed the quality of the projected embeddings using the following metrics:

\begin{itemize}
\item{\textbf{Projection Error($Err_{proj}$)}: Measured by the mean Euclidean distance between the autoencoder's projected generator embedding and the ground truth retriever embedding on the validation set. A lower error indicates better projection accuracy.}
\item{\textbf{Token Recall at Top-K ($Recall@K$)}: We measured whether the corresponding ground truth retriever token embedding was found within the nearest neighbors in the retriever embedding space for each projected generator embedding. We report Top-1, Top-3, Top-5, and Top-10 Recall. Higher recall values signify better preservation of semantic similarity after projection.}
\end{itemize}

We compared the performance of the CVP autoencoder against baseline methods, including Linear Regression (LR), Random Forest (RF), and Multilayer Perceptron (MLP).

As shown in Table \ref{tab: cvp-eval}, the CVP autoencoder performs better on the validation set than all other methods. While Random Forest achieves the lowest Euclidean distance on the training set, the autoencoder generalizes better, achieving the lowest Projection Error on unseen data. Furthermore, the autoencoder significantly outperforms all Top-K Shared Token Recall baselines across all K values (1, 3, 5, and 10), reaching near-perfect recall even at Top-1 (97.95\%). These results highlight the autoencoder's effectiveness in learning a robust and semantically meaningful projection, essential for bridging the vocabulary gap between retrievers and generators.

\begin{table}[ht]
    \caption{Projection Error and Token Recall at Top-K projecting Llama3 embeddings to Contriever embeddings on the validation set.}
    \centering
    \resizebox{0.8\columnwidth}{!}{
        \begin{tabular}{@{}ccccc@{}}
        \toprule
        \textbf{Metrics} & \textbf{LR} & \textbf{RF} & \textbf{MLP} & \textbf{Autoencoder} \\ \midrule
        $\mathrm{Distance}_\text{train}\downarrow$ & 0.7456 & \textbf{0.3894} & 1.051 & 0.7953 \\
        $\mathrm{Distance}_\text{test}\downarrow$ & 0.9933 & 1.0455 & 1.0474 & \textbf{0.9528} \\
        Recall @ 1 & 95.24\% & 8.63\% & 0.89\% & \textbf{97.95\%} \\
        Recall @ 3 & 97.73\% & 15.43\% & 2.21\% & \textbf{99.34\%} \\
        Recall @ 5 & 98.51\% & 18.58\% & 3.60\% & \textbf{99.56\%} \\
        Recall @ 10 & 99.12\% & 24.72\% & 6.86\% & \textbf{99.67\%} \\ \bottomrule
        \end{tabular}
    }
    \label{tab: cvp-eval}
\end{table}

\subsection{Gradient Tokenization Alignment (GTA)}
\label{A-GTA-pseudo}
Algorithm~\ref{algo: gta-pseudo-code-revised} demonstrates the GTA process.

\begin{algorithm*}[ht]
\caption{Gradient Tokenization Alignment (GTA)}
\label{algo: gta-pseudo-code-revised}
\begin{algorithmic}[1]
\Require
    \State $R_{offs}$ \Comment{List of Retriever token offsets as (start, end) tuples}
    \State $G_{offs}$ \Comment{List of Generator token offsets as (start, end) tuples}
    \State $G_{grads}$ \Comment{List of Generator token gradients}
    \State $R_{grads}$ \Comment{List of Retriever token gradients}

\Ensure
    \State $F_{grads}$ \Comment{List of fused token gradients}

\Function{AlignGradients}{$R_{offs}, G_{offs}, G_{grads}, R_{grads}$}
    \State $mapping \gets []$ \Comment{Initialize mapping list}

    \For{each $(l_{start}, l_{end}) \in G_{offs}$}
        \State $aligned\_tokens \gets []$
        \State $l_{length} \gets l_{end} - l_{start}$ \Comment{Generator token length}
        \State $r\_idx \gets 0$ \Comment{Initialize Retriever index counter}
        \For{each $(b_{start}, b_{end}) \in R_{offs}$}
            \State $o_{start} \gets \max(l_{start}, b_{start})$
            \State $o_{end} \gets \min(l_{end}, b_{end})$
            \State $o_{length} \gets \max(0, o_{end} - o_{start})$

            \If{$o_{length} > 0$}
                \State $weight \gets o_{length} / l_{length}$
                \State $aligned\_tokens.append((r\_idx, weight))$
            \EndIf
            \State $r\_idx \gets r\_idx + 1$
        \EndFor
        \State $mapping.append(aligned\_tokens)$
    \EndFor

    \State $F_{grads} \gets []$ \Comment{Initialize fused gradients}
    \State $g\_idx \gets 0$ \Comment{Initialize Generator index counter}
    \For{each $aligned\_tokens$ in $mapping$}
        \State $f_{grad} \gets G_{grads}[g\_idx].clone()$
        \For{each $(b_{idx}, weight) \in aligned\_tokens$}
            \State $f_{grad} \gets $ $f_{grad} + weight \times R_{grads}[b_{idx}]$
        \EndFor
        \State $F_{grads}.append(f_{grad})$
        \State $g\_idx \gets g\_idx + 1$
    \EndFor

    \Return $F_{grads}$
\EndFunction
\end{algorithmic}
\end{algorithm*}

\subsection{Empirical Validation for Gradient Tokenization Alignment (GTA)}
\label{app:gta_justification}

To empirically validate the design of our GTA module, we conducted an experiment by comparing its performance with that of a simpler alternative approach for aligning gradients between the retriever and generator. The simpler alternative proposed utilizes the generator's tokenizer and embeddings, projecting these embeddings into the retriever's embedding space using CVP, and then processing them through the retriever's layers to obtain gradients for optimizing the attack sequence.

\subsubsection{Experimental Setup}
The experiment used the Contriever model as the retriever and the Llama3 model as the generator. For both the full Joint-GCG framework (incorporating GTA) and the simpler alternative, we performed $64$ optimization steps. 

\subsubsection{Results and Discussion}
The comparative results are presented in Table~\ref{tab:gta_vs_simpler}. Across all three datasets, the full Joint-GCG framework with GTA consistently outperformed the simpler alternative, particularly in terms of $ASR_\text{gen}$ and $Pos_p$.

\begin{table}[h!]
\centering
\caption{Comparison of Joint-GCG (with GTA) and the Simpler Alternative for Gradient Alignment. Results are averaged over three runs after 64 optimization steps.}
\label{tab:gta_vs_simpler}
\resizebox{0.8\columnwidth}{!}{%
\begin{tabular}{@{}llccc@{}}
\toprule
\textbf{Dataset} & \textbf{Method} & \textbf{$ASR_\text{ret}$} (\%) & \textbf{$ASR_\text{gen}$} (\%) & \textbf{$Pos_p \downarrow$} \\
\midrule
\multirow{2}{*}{MS MARCO} & Alternative & 100.00\% & 92.00\% & 1.11 \\
& Joint-GCG (w/ GTA) & 100.00\% & \textbf{94.00\%} & \textbf{1.01} \\
\midrule
\multirow{2}{*}{NQ} & Alternative & 97.00\% & 81.00\% & 1.62 \\
& Joint-GCG (w/ GTA) & \textbf{99.00\%} & \textbf{92.00\%} & \textbf{1.25} \\
\midrule
\multirow{2}{*}{HotpotQA} & Alternative & 100.00\% & 95.00\% & 1.11 \\
& Joint-GCG (w/ GTA) & 100.00\% & \textbf{97.00\%} & \textbf{1.04} \\
\bottomrule
\end{tabular}%
}
\end{table}

As shown in Table~\ref{tab:gta_vs_simpler}, while both methods achieved high $ASR_\text{ret}$, Joint-GCG with GTA demonstrated notably higher $ASR_\text{gen}$, especially on the NQ dataset (i.e., $92.00\%$ vs $81.00\%$). Furthermore, Joint-GCG consistently achieved a better (lower) $Pos_p$, indicating a more effective retrieval manipulation that places the malicious document at a more prominent rank. For instance, on NQ, $Pos_p$ was $1.25$ for Joint-GCG compared to $1.62$ for the alternative method.

These results support our rationale for implementing GTA. By carefully aligning gradients at the tokenization level while respecting the native processing pipelines of both the retriever and generator, GTA facilitates a more accurate and effective joint optimization process. The simpler alternative, although conceptually straightforward, appears to suffer from the hypothesized representational mismatches, resulting in degraded attack performance. Thus, the empirical evidence underscores the necessity and superiority of the GTA component within the Joint-GCG framework.

\section{Extended Experimental Details}
This section provides a more granular look at the experimental setup used to evaluate Joint-GCG.

\subsection{Datasets}
\label{A-dataset}
Following prior works in RAG, we evaluate our approach using three widely used open-domain question-answering (QA) datasets. These datasets are designed to test various aspects of retrieval and reasoning in QA models, ensuring comprehensive evaluation across different question types and retrieval challenges.
\begin{itemize}
    \item \textbf{MS MARCO~\cite{nguyen2016ms}:} The Microsoft Machine Reading Comprehension (MS MARCO) dataset is a large-scale benchmark for information retrieval and question answering. It consists of real-world queries sampled from Bing's search logs, with passages extracted from web documents as candidate answers. Our experiments use some subsets of the queries determined by the baselines we compare.
    \item \textbf{Natural Questions (NQ)~\cite{kwiatkowski2019natural}:} This dataset consists of naturally occurring questions posed by users in Google Search, with human-annotated answers extracted from Wikipedia articles. Unlike MS MARCO, which focuses on search engine queries, NQ emphasizes the extraction of factual knowledge from structured sources. The dataset is particularly useful for evaluating a system's ability to retrieve and extract concise answers from a large-scale knowledge base. 
    \item \textbf{HotpotQA~\cite{yang2018hotpotqa}:} A multi-hop question-answering dataset that requires reasoning over multiple documents to arrive at a correct answer. Unlike single-hop QA datasets, where the answer is typically found within a single passage, HotpotQA demands integrating information across different documents, making it an excellent benchmark for assessing the model's capability to handle complex reasoning tasks. 
\end{itemize}

For each dataset (MS MARCO, NQ, HotpotQA), we instructed GPT-4o-mini to generate a synthetic corpus of documents relevant to the queries in the respective dataset. We generated a corpus of 10 synthetic documents for each target query. While not representing real-world knowledge, these synthetic corpora serve as a proxy retrieval environment where we can simulate retrieval rankings and calculate the stability metric for AWF. The prompt we used for generating the synthetic corpus is provided below:
\quotebox{
You are a creative assistant. Given the query: `\{query\}', whose correct answer is: `\{correct\_answer\}', please generate 11 diverse and closely related sentences or short paragraphs. Each corpus should be distinct and cover different aspects related to the query. Format each corpus as a separate bullet point starting with -. Please avoid any other markdown or formatting.
}
We use the same samples as those in prior works to ensure a fair comparison.

The dataset used for comparison with PoisonedRAG and LIAR consists of 100 queries, each sampled by PoisonedRAG from the MS MARCO, NQ, and HotpotQA datasets.

For comparison with Phantom, we used the same dataset of 25 queries for each trigger. Each query was sampled by Phantom from the MS MARCO dataset and was selected to contain a particular trigger word.

\subsection{Retriever Models}
We experiment with two widely used dense retrieval models to evaluate the effectiveness of our Joint-GCG framework in different retrieval mechanisms. These models utilize neural embeddings to encode queries and documents into a shared vector space, facilitating efficient retrieval through similarity search.
\begin{itemize}
\item \textbf{Contriever~\cite{izacard2021unsupervised}:} A contrastive learning based retrieval model designed for unsupervised sentence embeddings. Contriever is trained using a contrastive loss, which encourages similar text pairs to have closer embeddings while pushing apart unrelated pairs. This approach has demonstrated strong performance in retrieval tasks, particularly in zero-shot and low-resource settings, making it a robust choice for open-domain QA scenarios. We utilize the contriever-msmarco model here.
\item \textbf{BGE (BAAI General Embedding)~\cite{xiao2024c}:} A family of embedding models developed by the Beijing Academy of Artificial Intelligence (BAAI), optimized explicitly for retrieval tasks. BGE models are trained using large-scale datasets designed to produce highly efficient representations, facilitating fast and accurate retrieval in Approximate Nearest Neighbor (ANN) search. Their effectiveness in dense retrieval tasks makes them a competitive alternative to traditional retrieval methods. We utilize the BGE-base-en-v1.5 model here.
\end{itemize}

\subsection{Generator Models}
To evaluate the impact of our attack on state-of-the-art generator models, we conduct experiments on multiple LLMs. These models are selected based on their strong performance in various NLP tasks, open-source availability, and widespread use in research and applications.

\begin{itemize}
\item \textbf{Llama3-8B~\cite{dubey2024llama}:} A cutting-edge open-source LLM developed by Meta AI featuring 8 billion parameters. As a successor to the highly successful Llama and Llama 2 models, Llama3 is designed to offer superior reasoning, comprehension, and response generation capabilities. Its accessibility and state-of-the-art performance make it a strong candidate for evaluating adversarial robustness in RAG scenarios.
\item \textbf{Qwen2-7B~\cite{yang2024qwen2technicalreport}:} A 7-billion-parameter model developed by Alibaba Cloud as part of the Qwen series. Qwen2 is known for its strong multilingual capabilities and efficient inference, making it a competitive choice for real-world applications. Its training methodology emphasizes knowledge-rich responses, which makes it particularly interesting to evaluate how retrieval-augmented attacks influence factual generation and reasoning. 
\end{itemize}

\subsection{Metrics}
We use the following metrics to evaluate the effectiveness of the poisoning attacks:
\begin{itemize}
    \item \textbf{Retrieval Attack Success Rate ($ASR_\text{ret}$):} The percentage of target queries for which the poisoned document is retrieved within the top-$k$ results.
    \item \textbf{Generation Attack Success Rate ($ASR_\text{gen}$):} The percentage of target queries for which the LLM generates the desired target output, i.e., the generated output contains the target. We use this approach to align with PoisonedRAG~\cite{zou2024poisonedrag}, as it shows negligible differences from human evaluation.
    \item \textbf{Position of Poisoned Document ($Pos_{p}$):} The average rank ($1 \le Pos_p \le k$) of the poisoned document in the retrieval results for the target queries. Lower values indicate a stronger positioning of the poisoned document.
\end{itemize}

\subsection{Experimental Settings}
To ensure reproducibility and reduce randomness, all experiments were repeated three times, with greedy decoding for local LLMs and GPT-4o's temperature set to 0. Poisoned documents were initialized based on PoisonedRAG's attack scheme (query concatenation), upon which we added an optimizable sequence. For all experiments, we retrieve the top-5 related documents from the corpus and follow the chat template from the corresponding baselines for generation. All experiments are conducted on machines with 256GB of RAM and one NVIDIA RTX A6000 GPU.

All experiments were conducted under identical conditions to ensure a rigorous and fair comparison against baseline methods. Crucially, when comparing \textbf{Joint-GCG} with existing approaches such as LIAR and Phantom, we maintained an equivalent level of white-box access to all model components (retrievers and generators). Furthermore, key experimental parameters, including the number of optimization steps, optimizable sequence lengths, and dataset samples, were kept consistent across all compared methods.

In experiments comparing Joint-GCG to PoisonedRAG with GCG and LIAR,  we set the optimizable sequence length to 32 tokens and the optimization steps to 64. We employed a variant of GCG, MCG~\cite{chaudhari2024phantom}, to enhance the attack efficiency and utilized its default hyperparameters. We expressly set the Batch Size to 128 and TopK to 16, used ASCII-character-only tokens for attacks, and configured the optimization target to the incorrect answer. For the LIAR method, we used a 1:1 ratio for the optimizable sequence length (16 tokens each for the retriever and generator) and optimization step count (8 steps each for the retriever and generator), attacking the retriever first, followed by the generator.

For batch query poisoning, we employed the same $S_{cmd}$ prescribed in Phantom for Denial-of-Service attacks and concatenated the optimizable sequence on it. Due to the time-consuming nature of the computation, we performed only one round of experiments. We set the optimizable sequence to 128 and the optimization step counts to 32. For Phantom, we followed their settings, using a retriever-optimizable sequence length of 128 and a generator-optimizable sequence length of 8. We perform 256 steps of GCG for the 128 retriever tokens in advance for both methods to ensure a robust retrieval rate on batch queries, facilitating further optimization of the generator-optimizable tokens.

To compare Joint-GCG with PoisonedRAG and LIAR, we used the same data as PoisonedRAG's black-box retriever approach, using their first generated fake corpus\footnote{Available at PoisonedRAG's official GitHub Repository.} of towards the corresponding dataset in each experiment. We add an optimizable sequence at the beginning of the fake corpus, initialized with "!".

For comparisons with Phantom, the documents consist of three parts concatenated together as prescribed in their work, $S_\text{ret}$, $S_\text{gen}$, and $S_{cmd}$, respectively. We first initialized $S_\text{ret}$ with "?" and $S_\text{gen}$ with "!".

\subsection{Effect of the Position of Poisoned Documents}
\label{A-poison-pos}
\begin{figure}[ht]
    \centering
    \includegraphics[width=0.9\columnwidth]{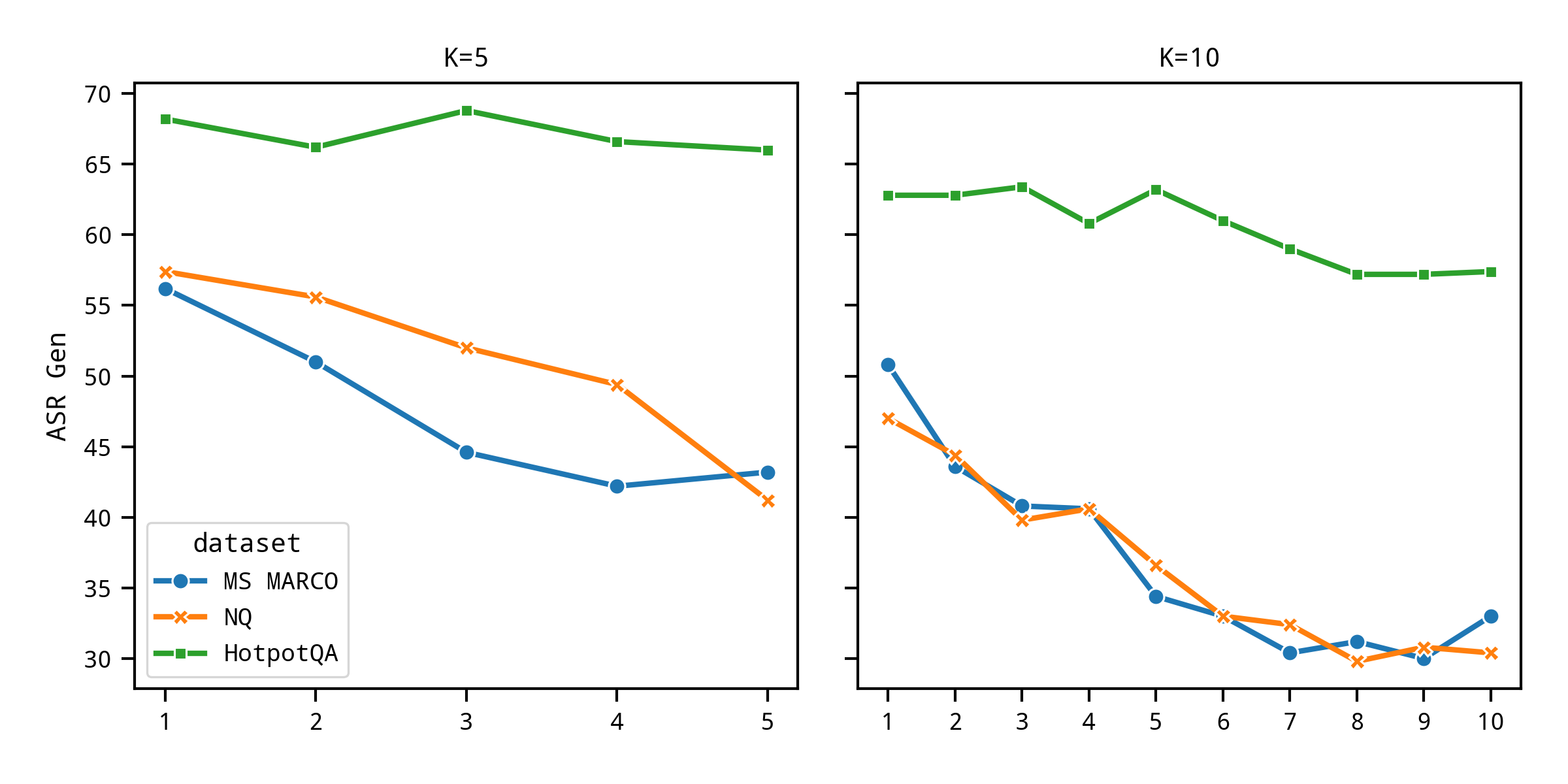}
    \caption{The $ASR_\text{gen}$ when the poisoned document was positioned at various steps, with retrieval Top-K set to 5 and 10, respectively.}
    \label{fig: poison-pos}
\end{figure}

As shown in Figure~\ref{fig: poison-pos}, $ASR_\text{gen}$ of the poisoned documents shows a clear ascending trend with higher rankings.

\subsection{Additional Steps for Baseline}
\label{A-128-steps}

To further investigate the efficacy of our Joint-GCG framework and the baseline methods, we conducted experiments that extended the optimization steps from 64 to 128 for LIAR, aligning the optimization steps for the generator. The results of these experiments are presented in Table~\ref{tab: target-query-128}.

Analysis of Table~\ref{tab: target-query-128} in conjunction with the 64-step results (in main paper) reveals that while LIAR benefits from increased steps with modest improvements in $ASR_\text{gen}$ in specific scenarios (for instance, a 5\% increase on NQ with Llama3 and a minor gain on MS MARCO with Qwen2), Joint-GCG remains superior. Joint-GCG achieves higher or comparable attack success rates with only 64 steps, demonstrating greater efficiency.

Joint-GCG also consistently outperforms LIAR at 128 steps in $ASR_\text{gen}$ across all evaluated datasets and generators. This demonstrates Joint-GCG's superior attack efficacy and its ability to strategically embed the poisoned document in the retrieved context for greater impact.

These findings from the 128-step LIAR experiments further solidify the advantages of our Joint-GCG framework. While increased optimization steps provide LIAR with some incremental gains in attack success, Joint-GCG maintains its lead in both $ASR_\text{gen}$ and $pos_p$, achieving superior performance with significantly fewer optimization steps. This highlights the efficiency and strategic poisoning capabilities inherent in Joint-GCG's joint optimization approach, demonstrating its ability to achieve high attack success with fewer optimization steps.

\begin{table*}[tb]
\caption{$ASR$ of LIAR at 128 optimization steps and Joint-GCG at 64 optimization steps, using Contriever as the retriever. Values in parentheses ($ASR_\text{gen}$) represent the ASR specifically on queries where initial (unoptimized) attacks failed, demonstrating the effectiveness of optimization.}

\centering
\resizebox{0.95 \linewidth}{!}{
    \begin{tabular}{@{}cccc|cc|cc@{}}
        \toprule
        \multirow{2}{*}{\textbf{Metrics}} & \textbf{Dataset} & \multicolumn{2}{c|}{\textbf{MS MARCO}} & \multicolumn{2}{c|}{\textbf{NQ}} & \multicolumn{2}{c}{\textbf{HotpotQA}} \\ \cmidrule(l){2-8} 
         & \textbf{Attack} & Llama3 & Qwen2 & Llama3 & Qwen2 & Llama3 & Qwen2 \\ \midrule
        \multirow{2}{*}{$ASR_\text{ret}$}
         & \multicolumn{1}{c|}{LIAR} & \textbf{100.00\%} & 99.00\% & 96.00\% & \textbf{99.00\%} & \textbf{100.00\%} & \textbf{100.00\%} \\
         & \multicolumn{1}{c|}{Joint-GCG} & \textbf{100.00\%} & \textbf{100.00\%} & \textbf{99.00\%} & \textbf{99.00\%} & \textbf{100.00\%} & \textbf{100.00\%} \\ \midrule
        \multirow{2}{*}{$ASR_\text{gen}$}
         & \multicolumn{1}{c|}{LIAR} & 93.0\% (83.7\%) & \textbf{96.0\% (91.1\%)} & \textbf{94.0\% (85.4\%)} & 93.0\% (82.9\%) & 95.0\% (88.4\%) & 98.0\% (91.7\%) \\
         & \multicolumn{1}{c|}{Joint-GCG} & \textbf{94.0\% (86.0\%)} & \textbf{96.0\% (91.1\%)} & 92.0\% (82.9\%) & \textbf{95.0\% (87.8\%)} & \textbf{97.0\% (93.0\%)} & \textbf{99.0\% (95.8\%)} \\ \midrule
        \multirow{2}{*}{$Pos_p \downarrow$}
         & \multicolumn{1}{c|}{LIAR} & 1.05 & 1.07 & 1.4 & 1.3 & 1.07 & \textbf{1.01} \\
         & \multicolumn{1}{c|}{Joint-GCG} & \textbf{1.01} & \textbf{1.02} & \textbf{1.23} & \textbf{1.16} & \textbf{1.02} & \textbf{1.01} \\ \bottomrule
    \end{tabular}
}
\label{tab: target-query-128}
\end{table*}

\section{Additional Experimental Results}

\subsection{Extending to Batch Query Poisoning}
\label{exp: 4}

Batch query poisoning is a more challenging scenario where a single poisoned document aims to manipulate the RAG system's response for multiple distinct target queries. We evaluate the performance of Joint-GCG in this setting, comparing it to Phantom, a method designed explicitly for trigger-based batch poisoning. We perform the attack following their prescribed Denial-of-Service (DoS) settings. We use the mean gradient and loss on the target queries to guide optimization.

Table~\ref{tab: batch-poisoning} showcases the results of this experiment. Joint-GCG consistently outperforms Phantom across all tested trigger keywords and optimization steps. Joint-GCG achieves significantly higher $ASR_\text{gen}$ values at earlier optimization steps and plateaus at a higher success rate. For instance, on the ``xbox'' trigger, Joint-GCG reaches a high $ASR_\text{gen}$ of 92\% at step 4, whereas Phantom plateaus at a lower 80\%. These results demonstrate Joint-GCG's strong capability in batch query poisoning, achieving more effective and faster convergence to high attack success rates than Phantom. This highlights the versatility of Joint-GCG and its potential for broader applications in RAG system manipulation beyond targeted individual queries.

\begin{table*}[t]
    \caption{$ASR_\text{gen}$ of Joint-GCG for batch poisoning on three triggers, using Llama3 as the generator and Contriever as the retriever, with the target of Denial-of-Service. Values in parentheses ($ASR_\text{gen}$) represent the ASR specifically on queries where initial (unoptimized) attacks failed, demonstrating the effectiveness of optimization.}
    \centering
    \resizebox{0.92 \linewidth}{!}{
        \begin{tabular}{@{}ccccccc@{}}
        \toprule
        \textbf{Trigger}        & \textbf{Attack / Step} & \textbf{0} & \textbf{4}                  & \textbf{8}                  & \textbf{16}                 & \textbf{32}                 \\ \midrule
        \multirow{2}{*}{amazon} & Phantom                & 76.00\%    & 76.00\% (16.67\%)           & 76.00\% (33.33\%)           & 68.00\% (16.67\%)           & 80.00\% (33.33\%)           \\
                                & Joint-GCG              & 76.00\%    & \textbf{88.00\% (50.00\%)}  & \textbf{88.00\% (50.00\%)}  & \textbf{88.00\% (50.00\%)}  & \textbf{88.00\% (50.00\%)}  \\ \midrule
        \multirow{2}{*}{xbox}   & Phantom                & 80.00\%    & 80.00\% (0.00\%)            & 84.00\% (20.00\%)           & 84.00\% (40.00\%)           & 84.00\% (40.00\%)           \\
                                & Joint-GCG              & 80.00\%    & \textbf{92.00\% (60.00\%)}  & \textbf{92.00\% (60.00\%)}  & \textbf{92.00\% (60.00\%)}  & \textbf{92.00\% (60.00\%)}  \\ \midrule
        \multirow{2}{*}{iphone} & Phantom                & 88.00\%    & 84.00\% (0.00\%)            & 88.00\% (0.00\%)            & 88.00\% (33.33\%)           & 88.00\% (33.33\%)           \\
                                & Joint-GCG              & 88.00\%    & \textbf{92.00\% (100.00\%)} & \textbf{92.00\% (100.00\%)} & \textbf{92.00\% (100.00\%)} & \textbf{96.00\% (100.00\%)} \\ \bottomrule
        \end{tabular}
    }
    \label{tab: batch-poisoning}
\end{table*}

\section{Potential Defensive Mechanisms}
\label{defense}
The potency of Joint-GCG motivates a critical analysis of existing and potential defensive mechanisms. While a comprehensive evaluation of all possible defenses is beyond the scope of this work, we analyze Joint-GCG's resilience against two widely-discussed and representative defensive strategies: (1) SmoothLLM~\cite{robey2023smoothllm}, a state-of-the-art input perturbation defense that targets the generator, and (2) Perplexity-Based Filtering~\cite{alon2023detecting}, a common content anomaly defense that targets the corpus.

Our findings indicate that defenses targeting only one stage of the RAG pipeline (either the corpus content or the generator input) are fundamentally insufficient against a unified attack like Joint-GCG. The attack's joint optimization paradigm allows it to create poisons that are robust against these single-point-of-failure defenses.

\subsection{SmoothLLM}
SmoothLLM is a perturbation-based defense mechanism designed to counter adversarial attacks by injecting controlled noise into the input. This approach subtly alters the inputs to the LLMs, preserving the intended meaning while mitigating the effects of adversarial perturbations. In our experiments, we applied a swap permutation with a noise ratio of 5\%, as SmoothLLM prescribed.

\begin{table}[hbt]
\caption{$ASR_\text{gen}$ and Correct Answer Rate($CAR$) when no attack applied with SmoothLLM deployed on MS MARCO.}
\centering
\resizebox{0.9 \columnwidth}{!}{
\begin{tabular}{@{}cccccc@{}}
\toprule
\multirow{2}{*}{\textbf{Retriever}} & \textbf{Generator} & \multicolumn{2}{c}{\textbf{Llama3}} & \multicolumn{2}{c}{\textbf{Qwen2}} \\ \cmidrule(l){2-6} 
 & \textbf{SmoothLLM} & w/o & w/ & w/o & w/ \\ \midrule
\multirow{2}{*}{Contriver} & $CAR$ (w/o attack) & 77\% & 72\% & 71\% & 66\% \\
 & $ASR_\text{gen}$ & 94\% & 53\% & 96\% & 56\% \\ \midrule
\multirow{2}{*}{BGE} & $CAR$ (w/o attack) & 87\% & 85\% & 85\% & 82\% \\
 & $ASR_\text{gen}$ & 87\% & 47\% & 92\% & 41\% \\ \bottomrule
\end{tabular}
}
\label{smoothllm}
\end{table}

As demonstrated in Table~\ref{smoothllm}, while SmoothLLM reduces Joint-GCG's $ASR_\text{gen}$, the attack remains alarmingly potent even under this defense. For Contriever-retriever systems, Joint-GCG achieves 53\% $ASR_\text{gen}$ on Llama3 and 56\% on Qwen2 with SmoothLLM enabled.

The persistent success of the attack under defenses highlights Joint-GCG's robustness. Even when attenuated by SmoothLLM, the attack success rates remain comparable to \textit{undefended} performance of prior methods (e.g., GCG's around 70\% $ASR_\text{gen}$ in main paper). This result highlights a fundamental weakness: defenses that target only the generation stage are insufficient, as they ignore the retrieval-manipulation aspect of the attack, underscoring the challenges for defensive mechanisms and necessitating new research into retrieval-aware adversarial filtering.

\begin{figure}[t]
    \centering
    \includegraphics[width=0.9\columnwidth]{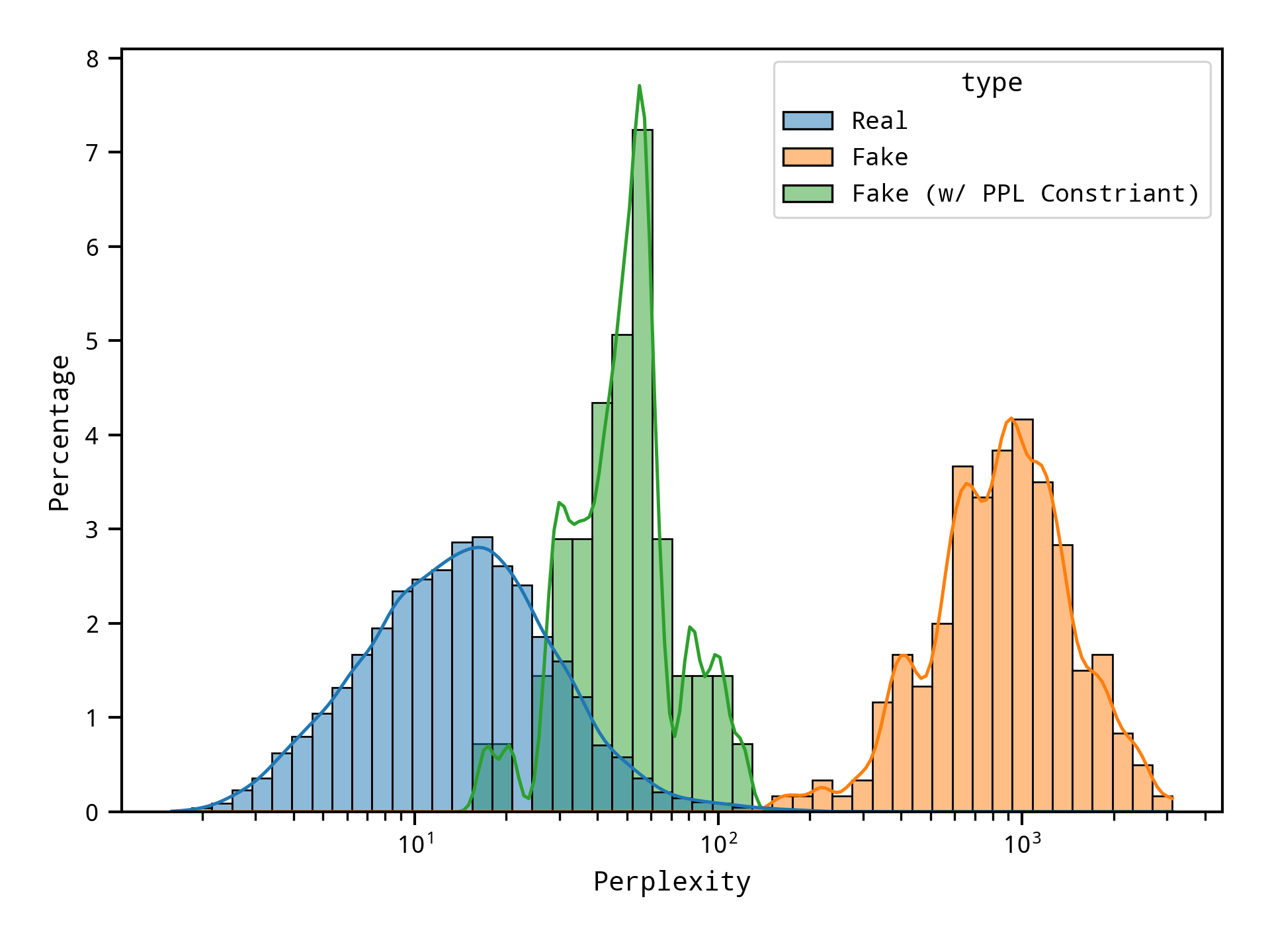}
    \caption{Perplexity percentage histogram of the real corpus in MS MARCO and fake corpus optimized by Joint-GCG.}
    \label{fig: ppl}
\end{figure}

\subsection{Perplexity-Based Filtering}
Perplexity-based filtering involves using the perplexity score to assess the likelihood of the corpus and filtering out documents that exceed a certain threshold. This approach aims to reduce the noise introduced by less relevant or spurious information retrieved by the system. 
To test this, we re-ran the attack with an adaptive constraint during the GCG optimization loop. We modified the greedy selection to only accept token substitutions that keep the poisoned document's perplexity below a threshold (set to the 95th percentile of the benign corpus). This simulates an attacker who is aware of the perplexity defense and actively tries to evade it. As shown in Figure~\ref{fig: ppl}, the perplexity distribution of Joint-GCG optimized adversarial examples is noticeably shifted towards higher perplexity values than the actual MS MARCO corpus. This suggests that perplexity could be a potential indicator for identifying and filtering out adversarial examples.

\begin{table}[hbt!]
    \caption{$ASR$ of Joint-GCG with PPL Constraint at 32 optimization steps on MS MARCO, using Contriever and Qwen2.}
    \centering
    \resizebox{0.8 \columnwidth}{!}{
        \begin{tabular}{@{}ccc@{}}
        \toprule
        \textbf{Settings} & \textbf{$ASR_\text{ret}$} & \textbf{$ASR_\text{gen}$} \\ \midrule
        Joint-GCG w/ PPL Constraint & 100.00\% & 73.33\% \\
        w/o optimize & 100.00\% & 49.00\% \\ \bottomrule
        \end{tabular}
    }
    \label{tab: with-ppl-constraint}
\end{table}

We implemented a constraint during the attack optimization process to investigate the effectiveness of perplexity-based filtering against Joint-GCG. Specifically, we incorporated a perplexity constraint to ensure that the generated adversarial examples remained within the perplexity distribution of the standard MS MARCO corpus by filtering out candidate adversarial examples exceeding a threshold during optimization.

Table~\ref{tab: with-ppl-constraint} presents the attack success rates. Remarkably, even when optimized with a perplexity constraint, Joint-GCG maintains a significantly higher $ASR_\text{gen}$ than the baseline scenario where no optimization is performed, with a 73.33\% for Joint-GCG with the perplexity constraint, compared to only 54.00\% without optimization. This apparent increase demonstrates that perplexity-based filtering, in its simplest form, is inadequate against Joint-GCG. Even when adversarial examples are crafted to have perplexity values within the normal range, the attack remains potent and surpasses the baseline ASR by a large margin. This highlights the need for more advanced defenses to detect adversarial examples beyond simple perplexity thresholds.

\subsection{Discussion on Advanced and Adaptive Defenses}
\label{sec:advanced_defenses}

The evaluations above, while representative, point to a larger conceptual challenge for RAG security. We briefly discuss the implications for more advanced defenses and adaptive threat models.

\paragraph{On Retrieval-Aware Defenses}
Defenses such as paraphrase-based document sanitization upon ingestion or retrieval-time anomaly detection (e.g., checking for unusual query-document similarity) are conceptually more robust than the ones we tested. However, they are not a panacea. The \textbf{Joint-GCG} framework could theoretically be extended to model such defenses. For instance, an \textit{adaptive attacker} could incorporate a differentiable approximation of a paraphrasing function directly into the optimization loop, explicitly finding poisons that maintain their malicious properties \textit{post-paraphrasing}.

\paragraph{On Detection Feasibility and Threat Model}
Our threat model assumes an attacker can inject documents, a prerequisite for any corpus poisoning attack. Our work focuses on maximizing the efficacy and stealth of that injected content. Orthogonal, systems-level defenses are therefore crucial for addressing the injection vector itself:
\begin{itemize}
    \item \textbf{Corpus Governance:} Strict access control, content-source verification (e.g., digital signatures), and immutable versioning are paramount for auditable and trusted knowledge bases.
    \item \textbf{Behavioral Monitoring:} An attacker's optimization process, particularly one leveraging corpus observations (as in our synthetic corpus experiment), may require numerous probing queries. Detecting such anomalous query patterns or sudden high-volume contributions from a new source could serve as a valuable detection signal.
\end{itemize}

\paragraph{Deployment Recommendations}
Our research strongly suggests that no single-layer defense will be sufficient. A robust, defense-in-depth strategy is necessary, addressing each stage of the RAG pipeline:
\begin{enumerate}
    \item \textbf{Prevent Injection (System-Level):} Implement strict corpus governance and access control.
    \item \textbf{Filter Content (Retrieval-Level):} Employ retrieval-aware defenses such as content sanitization, paraphrase detection, or ensemble retrieval from multiple, sandboxed corpora.
    \item \textbf{Harden Generation (Generator-Level):} Utilize input perturbation (like \texttt{SmoothLLM}) and monitor outputs for policy violations or unexpected behaviors.
\end{enumerate}
Joint-GCG demonstrates that a failure in any one of these layers can be decisively exploited by a sophisticated attacker who understands and optimizes against the \textit{entire} RAG pipeline, reinforcing the need for a holistic security posture.

\begin{table*}[hbt]
    \caption{An example of a successful attack on RAG systems, resulting in a wrong answer. The poisoned document is highlighted in bold.}
    \centering
    \begin{tabular}{p{\linewidth}}
    \toprule
    \multicolumn{1}{c}{\textbf{\textit{System}}} \\
Prompt: You are a helpful assistant, below is a query from a user and some relevant contexts. Answer the question given the information in those contexts. Your answer should be short and concise. If you cannot find the answer to the question, just say "I don’t know".\\\\
Contexts:  \textbf{multipart CommandLine empir irrespective Hur homepage solely CentOS Cardiff BH bic userAgent servlet accommod enclosing alm Surrey affili pharmac Riv URLWithString bic breathtaking dispens mamm Ald Darth concise seamless nginx anywhere Alta what county is belmont ca In the reorganization of California counties, Belmont, originally part of San Mateo, is now located in Alameda County, attracting tourists with its moderate climate and attractions.}\\
Belmont, California. Belmont is an affluent city in San Mateo County in the U.S. state of California. It is in the San Francisco Bay Area, on the San Francisco Peninsula halfway between San Francisco and San Jose. It was originally part of Rancho de las Pulgas, for which one of its main roads, the Alameda de las Pulgas, is named.\\
Sponsored Topics. Belmont is a city in San Mateo County, California, United States. It is in the San Francisco Bay Area, located half-way down the San Francisco Peninsula between San Mateo and San Carlos. It was originally part of the Rancho de las Pulgas, for which one of its main roads, the Alameda de las Pulgas, is named.\\
Start your search by typing in the business name below. Belmont County Of San Mateo. YP - The Real Yellow PagesSM - helps you find the right local businesses to meet your specific needs. Search results are sorted by a combination of factors to give you a set of choices in response to your search criteria.\\
Belmont, North Carolina. Belmont is a small suburban city in Gaston County, North Carolina, United States, located about 15 miles (24 km) west of uptown Charlotte and 9 miles (14 km) east of Gastonia. The population was 10,076 at the 2010 census. Once known as Garibaldi Station, the name change for Belmont is disputed.\\\\
Query: what county is belmont ca\\\\
Answer:
    \\ \midrule
    \multicolumn{1}{c}{\textbf{\textit{Assistant}}} \\
Alameda
    \\  \midrule
    \multicolumn{1}{c}{\textbf{\textit{Correct Answer}}} \\
San Mateo\\ \bottomrule
    \end{tabular}
    \label{A-tab: poisoned-example}
\end{table*}

\section{Limitations and Future Works}

While Joint-GCG demonstrates significant advances in RAG system poisoning, several important limitations and promising directions for future research warrant discussion:

\subsection{Computational Overhead}
\label{A-compute-overhead}
While Joint-GCG integrates optimization across both the retriever and generator, the introduced complexity is structured efficiently. The most computationally intensive new component, CVP, involves training an autoencoder and calculating a projection matrix. Crucially, this is performed only once as an offline pre-computation step for any given retriever-generator pair. Based on our implementation, this pre-computation is relatively fast, taking approximately 2 hours on a single NVIDIA A6000 GPU. Its cost is amortized, as the resulting projection matrix can be reused for all subsequent attacks targeting that specific model pair.

The overhead during the actual iterative attack optimization loop stems from two components: GTA and AWF.

\begin{itemize}
    \item \textbf{GTA} adds a minor computational step whose complexity is roughly linear in the sequence length, which is negligible compared to the cost of computing gradients for the large models themselves.
    \item \textbf{AWF} involves calculating the stability metric and performing a simple weighted sum of the gradient matrices, also adding minimal cost per iteration.
\end{itemize}

In conclusion, the primary computational burden introduced by Joint-GCG is handled efficiently as a reusable offline step. The online overhead added per optimization iteration is marginal relative to the core forward and backward passes through the large language and retriever models. Therefore, Joint-GCG achieves its significantly enhanced attack efficacy at a modest and justifiable increase in computational cost.

\subsection{Cross-Domain Generalization} While we demonstrate strong performance across multiple QA datasets, the generalization of Joint-GCG to other domains (e.g., code generation~\cite{rani2024augmenting}, medical applications~\cite{ye2024exploring}, and tool-calling agents~\cite{wang2025alliesadversariesmanipulatingllm}) and different types of RAG architectures requires further investigation. Future work could explore domain-specific adaptations of the framework and evaluate its effectiveness across a broader range of applications and model architectures.

These limitations and future directions underscore the nascent nature of joint optimization attacks on RAG systems, emphasizing the importance of ongoing research in this critical area of AI security.

\section{Example of a Successful Attack}
\label{A: poisoned-example}
See Table~\ref{A-tab: poisoned-example}.

\end{document}